\documentclass[aip,jcp,preprint,prb,floats,aps,amsmath,amssymb,citeautoscript]{revtex4}

\usepackage{graphicx}
\usepackage{bm}
\usepackage{lscape}
\usepackage{amsmath}
\usepackage{braket}
\usepackage{longtable}
\usepackage{array}
\usepackage{natbib}
\usepackage{xcolor} 
\usepackage{epstopdf}

\begin{document}

\title{The Rotation-Vibration Structure of the SO$_2$ $\tilde{\text{C}}$$^1$B$_2$ State Explained by a New Internal Coordinate Force Field}
\author{Jun Jiang, G. Barratt Park, and Robert W. Field\footnote{Corresponding author. Email: rwfield@mit.edu}}
\affiliation{Department of Chemistry, Massachusetts Institute of Technology, Cambridge, Massachusetts 02139, USA}
\date{\today}

\begin{abstract}
A new quartic force field for the SO$_2$ $\tilde{\text{C}}$$^1$B$_2$ state has been derived, based on high resolution data from S$^{16}$O$_2$ and S$^{18}$O$_2$. Included are eight $b_2$ symmetry vibrational levels of S$^{16}$O$_2$ reported in the first paper of this series [G. B. Park, \textit{et al.}, J.\ Chem.\ Phys.\ \textbf{144}, 144311 (2016)]. Many of the experimental observables not included in the fit, such as the Franck-Condon intensities and the Coriolis-perturbed effective $C$ rotational constants of highly anharmonic $\tilde{\text{C}}$ state vibrational levels, are well reproduced using our force field. Because the two stretching modes of the $\tilde{\text{C}}$ state are strongly coupled via Fermi-133 interaction, the vibrational structure of the $\tilde{\text{C}}$ state is analyzed in a Fermi-system basis set, constructed explicitly in this work via $\textit{partial}$ $\textit{diagonalization}$ of the vibrational Hamiltonian. The physical significance of the Fermi-system basis is discussed in terms of semiclassical dynamics, based on study of Fermi-resonance systems by Kellman and coworkers [M. E. Kellman and L. Xiao, J. Chem. Phys. \textbf{93}, 5821 (1990)]. By diagonalizing the vibrational Hamiltonian in the Fermi-system basis, the vibrational characters of all vibrational levels can be determined unambiguously. It is shown that the bending mode cannot be treated separately from the coupled stretching modes, particularly at vibrational energies of more than 2000 cm$^{-1}$. Based on our force field, the structure of the Coriolis interactions in the $\tilde{\text{C}}$ state of SO$_2$ is also discussed. We identify the origin of the alternating patterns in the effective $C$ rotational constants of levels in the vibrational progressions of the symmetry-breaking mode, $\nu_\beta$ (which correlates with the antisymmetric stretching mode in our assignment scheme).  

\end{abstract}

\maketitle

\section{Introduction}
\label{sec:intro}

Pioneering work by Hoy and Brand~\cite{{Brand1978},{Brand1976}}, based on earlier analysis by Coon and co-workers~\cite{{Coon1966},{Jones1973}}, established the presence of a double-well structure in the anti-symmetric stretching coordinate on the potential energy surface (PES) of the SO$_2$ $\tilde{\text{C}}$$^1$B$_2$ state. That is, the minimum geometry of the $\tilde{\text{C}}$ state has nonequivalent S-O bond lengths. A strong Fermi interaction between the symmetric and anti-symmetric stretching modes was inferred from the anomalously intense transition into the first overtone of the anti-symmetric stretching mode. Using a reduced-dimension model that excludes terms containing $q_2$, Hoy and Brand~\cite{Brand1978} derived an internal coordinate force field of the $\tilde{\text{C}}$ state. 

The presence of a double well in the $q_3$ coordinate and a strong Fermi interaction between $\nu_1$ and $\nu_3$ was validated in subsequent studies of the $\tilde{\text{C}}$ state~\cite{{Johann1977},{Yamanouchi1995},{Sako1998},{Bludsky2000},{Xie1999},{Nachtigall1999},{Ran2007},{Katagiri1997},{Ebata1988}}. In a series of papers published in the late 1990s~\cite{{Yamanouchi1995},{Sako1998}}, Yamanouchi \textit{et al.}\ experimentally determined an additional 33 vibrational term values of $a_1$ vibrational symmetry levels~\cite{Yamanouchi1995}. Normal-mode assignments were given for some of the observed levels in their first paper~\cite{Yamanouchi1995}, although those assignments were later found to be inaccurate, given the large Fermi interactions. In the second paper~\cite{Sako1998}, Sako \textit{et al.}\ inspected the shapes of the $\tilde{\text{C}}$ state vibrational wavefunctions (obtained from the derived normal-mode force field) in the $q_1-q_3$ plane after integrating the 3D wavefunctions along the $q_2$ coordinate. The nodal patterns of the integrated wavefunctions were found to be characteristic of Fermi resonance systems~\cite{{Kellman1986},{Kellman1988},{Kellman1990cat},{Kellman1990new},{Kellman1990}}. The vibrational levels were assigned by a generalized vibrational assignment scheme, based on visual inspection of the nodal patterns. In Sako's work, the stretch-bend interactions were assumed to be less important than the stretch-stretch interactions for the $\tilde{\text{C}}$ state vibrational levels (and their assignments), and the effects from the stretch-bend interactions were effectively averaged out after the 3D wavefunctions were integrated along $q_2$. However, given that some of the derived stretch-bend interaction constants have non-negligible magnitudes (e.g. $\phi_{233}=-44$ cm$^{-1}$)~\cite{Sako1998}, the assumption of near-complete separability of $\nu_2$ from the other two strongly interacting modes is expected to break down for levels with higher quanta of excitation, although Sako et al. did not discuss when and how the breakdown occurs.

In addition to extensive anharmonic interactions, the majority of the $\tilde{\text{C}}$ state vibrational levels exhibit $c$-axis Coriolis perturbations~\cite{{Yamanouchi1995},{Park2015SO2},{Johann1977},{Brand1978},{Brand1976}}. Since the effects of Coriolis interactions are sensitive to the energy spacings between levels, a double-well structure on the PES, which gives rise to vibrational level-staggerings, is expected to cause staggering-related anomalies in the rotational structure of the $\tilde{\text{C}}$ state, especially in the vibrational levels that involve the symmetry-breaking mode. However, rotational information for levels with odd quanta of excitation in the symmetry-breaking mode had not been available until our recent direct observations of $b_2$ vibrational symmetry levels, reported in the first paper of this series~\cite{Park2015SO2} and in Ref ~\onlinecite{Park2015}. With those crucial pieces of information on the rotational structure of the $\tilde{\text{C}}$ state, we can now validate and interpret the rotational anomalies caused by the double-well structure on the PES. However, correct identification and detailed understanding of the anomalies require knowledge of the molecular force field. 

With direct high resolution measurements of the first eight $b_2$ symmetry vibrational levels~\cite{Park2015SO2}, and an additional $b_2$ level at 2754 cm$^{-1}$~\cite{Park2015}, we can now determine a more physical and accurate force field for the SO$_2$ $\tilde{\text{C}}$$^1$B$_2$ state. The accuracy of the force field of Hoy and Brand is limited by the reduced-dimension nature of their fit, and the fact that the $\nu_3$ fundamental level, the position of which was estimated from the inferred position of the (0,1,1) level~\cite{Brand1978}, was the only $b_2$ symmetry level included in the fit. Somewhat fortuitously, their estimated $\nu_3$ fundamental frequency was extremely accurate~\cite{Park2015SO2}. Yamanouchi et al.~\cite{{Yamanouchi1995},{Sako1998}} extended the Hamiltonian of Hoy and Brand to three dimensions. However, all the parameters associated with the double-well structure were fixed to the values derived from the Hoy and Brand 2D fit, and no $b_2$ symmetry levels were included in the fit. In addition, none of the available rotational and isotopologue information was used as inputs to the fit. 

In this paper, we derive an internal coordinate force field of the $\tilde{\text{C}}$ state of SO$_2$, which incorporates vibrational and rotational information of two isotopologues, $^{32}$S$^{16}$O$_2$ and $^{32}$S$^{18}$O$_2$. Our methods for calculating and fitting rovibrational information of the $\tilde{\text{C}}$ state of SO$_2$, e.g. vibrational band-origins, rotational constants, and Coriolis matrix elements, are described in Section~\ref{sec:methods}. In Section~\ref{sec:2D fit}, we present the result from a reduced-dimension fit, similar to the fit model developed by Hoy and Brand~\cite{Brand1978}. The reduced-dimension model provides insight into the unique effects of stretch-stretch Fermi resonance on the vibrational dynamics in the $\tilde{\text{C}}$ state. A scheme for assigning the 2D wavefunctions of the $\tilde{\text{C}}$ state is also discussed, based on semiclassical work by Kellman and co-workers~\cite{{Kellman1986},{Kellman1988},{Kellman1990cat},{Kellman1990new},{Kellman1990}}. In Section~\ref{sec:3D fit}, our 3D internal coordinate force field is reported. We demonstrate the accuracy and predictive power of our 3D force field by comparing the values of the experimental observables that are not directly used as inputs to our fit to the calculated values from our force field. In particular, the Franck-Condon (fc) intensities and the strongly perturbed effective $C$ rotational constants of highly anharmonic $\tilde{\text{C}}$ state vibrational levels are well reproduced. A two-step diagonalization procedure of the vibrational Hamiltonian is developed in Section~\ref{subsec:vib_assign}. The two-step diagonalization allows assignments and characterization of an unprecedented number of $\tilde{\text{C}}$ state vibrational levels in a new Fermi-system basis (the Kellman basis), constructed explicitly in this work via partial diagonalization of the Hamiltonian. We investigate the breakdown of the separability of $\nu_2$ from the other two strongly Fermi-interacting modes for levels $>$2000 cm$^{-1}$ above the $\tilde{\text{C}}$ state zero-point level. Based on our force field, the structure of the Coriolis interactions in the $\tilde{\text{C}}$ state is discussed in Section~\ref{subsec:Coriolis}. We are able to identify and explain the alternating patterns in the effective $C$ rotational constants for levels in the vibrational progressions of the symmetry-breaking mode.

\section{Methods}
\label{sec:methods}
A vibrational Hamiltonian of the following form is used to fit the vibrational band origins of the SO$_2$ $\tilde{\text{C}}$ state:
\begin{multline}
H/hc=\frac{1}{2} \omega_1 (q_1^2+p_1^2)+\frac{1}{2} \omega_2 (q_2^2+p_2^2)+\frac{1}{2} \omega_3 (q_3^2+p_3^2)\\
+\frac{1}{6} \phi_{111}q_1^3+\frac{1}{2}\phi_{112}q_1^2q_2+\frac{1}{2}\phi_{122}q_1q_2^2+\frac{1}{2}\phi_{133}q_1q_3^2+\frac{1}{2}\phi_{233}q_2q_3^2+\frac{1}{6}\phi_{222}q_2^3\\
+\frac{1}{24}\phi_{1111}q_1^4+\frac{1}{4}\phi_{1122}q_1^2q_2^2+\frac{1}{4}\phi_{1133}q_1^2q_3^2+\frac{1}{24}\phi_{2222}q_2^4+\frac{1}{4}\phi_{2233}q_2^2q_3^2+\frac{1}{24}\phi_{3333}q_3^4\\
+\frac{1}{6}\phi_{1222}q_1q_2^3+\frac{1}{6}\phi_{1112}q_1^3q_2+\frac{1}{2}\phi_{1233}q_1q_2q_3^2+C_1\exp{(\text{-}C_2q_3^2)},
\label{eq:Hamiltonian}
\end{multline}
where 
\begin{equation}
C_1=\frac{b\exp{(\rho)}}{\exp{(\rho)}-\rho-1},
\label{eq:C1}
\end{equation}
\begin{equation}
C_2=\omega_3[\exp{(\rho)}-\rho-1]/2b,
\label{eq:C2}
\end{equation}
and the $q$'s and $p$'s are dimensionless normal-mode coordinates and the conjugate momenta, respectively. Eq.~(\ref{eq:Hamiltonian}) is an expansion around C$_{2v}$ geometry through quartic terms of the pure vibrational part of the molecular Hamiltonian, with a Gaussian hump in the antisymmetric stretch $q_3$ direction to account for non-equivalent S-O bond-lengths. The Gaussian hump in the PES is defined by three parameters, $b$, $\omega_3$, and $\rho$~\cite{Coon1966}. The parameter, $b$, characterizes the height of the barrier along the $q_3$ direction (with $q_1=q_2=0$), and the parameter, $\rho$, gives information about the curvatures at the two minima of the PES along $q_3$. More detailed discussions regarding the parameters involved in the Gaussian hump can be found in Ref ~\onlinecite{Coon1966}. Through quartic terms, there are a total of eighteen normal-mode force constants. The eigenvalues and eigenvectors of the vibrational Hamiltonian are obtained from diagonalization of the Hamiltonian in the normal-mode basis. Details regarding the construction of the Hamiltonian matrices are given in Appendix~\ref{app:matrix}.

Since normal-mode coordinates depend on the atomic masses, the normal-mode force constants, $\omega$'s and $\phi$'s, are isotopologue-dependent~\cite{Papousek}. However, the observed vibrational levels of both $^{32}$S$^{16}$O$_2$ and $^{32}$S$^{18}$O$_2$ can be fit using $one$ set of isotopologue-independent internal force constants (the superscript on $^{32}$S will be dropped from here on). A transformation from normal-mode force constants to internal force constants is employed. This nonlinear transformation was worked out by Hoy, Mills, and Strey~\cite{Hoy1972}, and is summarized in Ref ~\onlinecite{Papousek}. We therefore omit discussion of details of the transformation.

Given that the $q_3$ coordinate is in a symmetry species of its own, the isotope dependence of the two remaining parameters defining the Gaussian hump, $b$ and $\rho$, can be determined by considering the 1D cross-section of the PES at $q_1=q_2=0$. We constrain $b$, which characterizes the barrier height along this cross-section, to be isotopologue-independent. The saddle point energy, $C_1$, should also remain unchanged with isotopic substitution, so we also constrain $\rho$ to be isotopologue-independent. 

Another benefit of using internal force constants is that they enable us to calculate and incorporate rotational information -- such as the rotational constants, centrifugal distortion coefficients, and Coriolis matrix elements -- into the fit. However, the very strong $q_3$ anharmonicity and the large Fermi-133 resonance interaction necessitate a special treatment of the rotational information. We have adopted the treatment of Hoy and Brand for the quartic centrifugal distortion constants of the zero-point vibrational level of both isotopologues~\cite{Brand1978}. The treatments of the rotational constants and the Coriolis matrix elements are summarized below, with additional information included in the Supplementary Material.\cite{supplementary}

The rotational constant, $\mathcal{R}_v$, of a vibrational level of the $\tilde{\text{C}}$ state of SO$_2$ is calculated by
\begin{equation}
\mathcal{R}_v=\sum_{n} c_{n,v}^{2}\mathcal{R}_{n},
\label{eq:rc}
\end{equation}
where $\mathcal{R}_{n}$ is the rotational constant of a normal-mode basis state, and $c_{n,v}$ is the coefficient of that basis state in the eigenvector that results from diagonalizing the Hamiltonian (Eq.~(\ref{eq:Hamiltonian})). Note that the symbol, $\mathcal{R}$, is used to generalize the notations for $A$, $B$, and $C$ rotational constants. The rotational constant of a normal-mode basis state is given by the conventional expression
\begin{equation}
\mathcal{R}_n=\mathcal{R}_e-\sum_{i} \alpha^\mathcal{R}_i (v_i+1/2),
\label{eq:rv}
\end{equation}
where $\mathcal{R}_e$ is the rotational constant at the equilibrium geometry, $v_i$ is the number of quanta in a specific normal-mode, and $\alpha_i$ is the rotation-vibration constant for that mode. The rotation-vibration constants are functions of the cubic normal-mode force constants. Some of the rotation-vibration constants also contain information about the Coriolis interactions in the molecule. For those reasons, the experimentally derived $\alpha^\mathcal{R}_i$ constants provide constraints on the derived force field. The general forms of rotation-vibration constants (see Supplementary Material) are derived by Mills via perturbation theory~\cite{{Mills1972},{Mills1974}}. Specific forms applicable to SO$_2$ are also available in Ref ~\onlinecite{Morino1964}. One advantage provided by Eq.~(\ref{eq:rc}) is that the rotational constants of a vibrational eigenstate can be calculated independent of the vibrational assignment of the eigenstate, which can be ambiguous due to anharmonic interactions, even in our assignment scheme discussed in Section~\ref{subsec:vib_assign}. The effects from the double-well and the resonant interactions on the rotational constants of a vibrational eigenstate are contained in the basis state expansion coefficients of the eigenstate.

In the $\tilde{\text{C}}$ state of SO$_2$, the $C$ rotational constants are strongly perturbed by $c$-axis Coriolis interactions~\cite{{Brand1978},{Brand1976},{Johann1977},{Yamanouchi1995}}. The $A$ and $B$ rotational constants, however, are unaffected by Coriolis interactions up to second-order of perturbation theory. In cases where non-degenerate perturbation theory is valid, Coriolis contributions to rotational constants can be included in the $\alpha_i$ parameter, and Eq.~(\ref{eq:rc}) can be used to calculate rotational constants of the perturbed levels. However, this approach fails for all of the $C$ rotational constants of  the $\tilde{\text{C}}$ state of SO$_2$ (even those which are not severely perturbed by Coriolis interactions), due to the presence of the double-well on the PES. Using second-order perturbation theory, the Coriolis contributions ($C_{Cor}$) to the $C$ constant of a vibrational level of the $\tilde{\text{C}}$ state must instead be calculated by the general expression
\begin{equation}
C_{Cor}=\sum_{v'} \bra{v}h\ket {v'}\bra{v'}h\ket {v}/(E_{v}-E_{v'}),
\label{eq:Cor}
\end{equation}
where
\begin{equation}
h=2C_e\zeta_{23}^c\left [\sqrt{\frac{\omega_3}{\omega_2}}q_2p_3-\sqrt{\frac{\omega_2}{\omega_3}} q_3p_2 \right ].
\label{eq:h21}
\end{equation} 
In Eq.~(\ref{eq:Cor}), the eigenvalues and eigenvectors of the vibrational Hamiltonian in Eq.~(\ref{eq:Hamiltonian}) are used for both $\ket{v}$ and the intermediate state $\ket{v'}$'s. The operator, $h$, defined in Eq.~(\ref{eq:h21}), is part of the Coriolis term, $hJ_c$, in the molecular Hamiltonian~\cite{Mills1972}. If we exclude the Coriolis contributions to the rotation-vibration constants, the rotational constants calculated using Eqs.~(\ref{eq:rc})-(\ref{eq:rv}) correspond to the Coriolis-deperturbed rotational constants from the experiments. The perturbed value of the $C$ rotational constant of a vibrational level, $C_p$, is then the sum of the deperturbed $C$ constant, $C_{dp}$, and $C_{Cor}$, or $C_p=C_{dp}+C_{Cor}$.

In our fit, the $A$ and $B$ constants of the three fundamental levels, as well as those of the zero-point vibrational level, are included. This helps to ensure a physical determination of both the equilibrium geometry and the cubic force parameters. The experimental $C$ constants are used to validate the goodness of our internal coordinate force field by comparing their values to the $C$ constants calculated from our force field. 

Coriolis matrix elements between $a_1$ and $b_2$ symmetry vibrational levels of the $\tilde{\text{C}}$ state of SO$_2$ derived from fits to the observed energy levels~\cite{Park2015SO2} are not included directly in the fit. However, the Coriolis matrix elements can be calculated using the derived force field and the calculated values may be compared to the experimentally derived values. The Coriolis matrix element, $t_1$, between vibrational eigenstates $\ket{v}$ and $\ket{v'}$ is defined as
\begin{equation}
t_1\equiv \frac{1}{2}\bra{v}h\ket{v'}.
\label{eq:me}
\end{equation}
The eigenvectors, $\ket{v}$ and $\ket{v'}$, are calculated from the force field. In this work, Coriolis interactions between modes $\nu_1$ and $\nu_3$ are neglected, since the $\nu_1$ and $\nu_3$ frequencies are very different, and $\zeta_{13}^c$ is about three times smaller than $\zeta_{23}^c$.

To summarize our fit procedure, internal force constants through quartic terms expanded about the C$_{2v}$ geometry are used as parameters in our force field fit of the $\tilde{\text{C}}$ state of SO$_2$. From those parameters, we derive the isotopologue-specific normal-mode force constants used in Eq.~(\ref{eq:Hamiltonian}) for both S$^{16}$O$_2$ and S$^{18}$O$_2$. The Hamiltonian (Eq.~(\ref{eq:Hamiltonian})) for each isotopologue is diagonalized to obtain eigenvalues and eigenvectors. The eigenvalues are fit to the observed vibrational band origins. The isotope shift of the zero-point level between the two isotopologues is also calculated and fitted to observed value. From the internal coordinate force field and the equilibrium geometry, we derive the rotational constants of each vibrational eigenstate, using Eqs.~(\ref{eq:rc})-(\ref{eq:rv}). The derived $A$ and $B$ rotational constants of the fundamental levels, and those of the zero-point vibrational level, are fitted to the experimental values for both isotopologues. All five quartic centrifugal distortion coefficients of the zero-point vibrational level for each isotopologue are also calculated from the internal force field and included in the fit. A nonlinear least-square fit is carried out to derive the internal coordinate force constants. 

Because we include vibrational and rotational information of two isotopologues, the number of data in our fit far exceeds that of all previous force field fits on the $\tilde{\text{C}}$ state of SO$_2$~\cite{{Yamanouchi1995},{Sako1998},{Xie1999}}. We have therefore chosen to use 23 fit parameters (one of which is constrained), compared with 17 in Yamanouchi's normal-mode force constants fit.~\cite{Sako1998} ~\footnote{Note that there are 15 parameters in total in their vibrational Hamiltonian. For their Franck-Condon calculation, they also need equilibrium geometry, which are presumably obtained from Hoy and Brand's 2D fit. This means effectively, they are using 17 parameters in their fit and calculation.}

To validate our internal coordinate force field, the $C$ rotational constants and Coriolis matrix elements are calculated from the force field. The calculated values are compared to the experimentally determined values. In addition, Franck-Condon factors between the $\tilde{\text{X}}$ state zero-point vibrational level and the $\tilde{\text{C}}$ state vibrational levels are calculated. The vibrational overlap integrals with the harmonic basis states of the $\tilde{\text{C}}$ state are calculated by the method of Sharp and Rosenstock~\cite{Sharp1964}, using the $\tilde{\text{X}}$ state geometry and quadratic force field parameters from Ref ~\onlinecite{Mills1974}. Anharmonic Franck-Condon factors for the $\tilde{\text{C}}$ state vibrational eigenstates are then calculated from the harmonic basis expansion coefficients. 

\section{Reduced-dimension fit}
\label{sec:2D fit}

\begin{table}[t]
\caption{Internal force constants of the $\tilde{\text{C}}$ state of SO$_2$ obtained from a two-dimensional fit, and normal-mode force constants of S$^{16}$O$_2$. Internal and normal-mode force constants derived by Hoy and Brand~\cite{Brand1978} are included for comparison. All normal-mode force constants have units of cm$^{-1}$. The mdyn-$\text{\AA}$ unit system is used for the internal force constants, e.g. [$f_{rr}$]=mdyn/$\text{\AA}$,  [$f_{r\theta}$]=mdyn, [$f_{rrr}$]=mdyn/$\text{\AA}^2$, etc., where 1 mdyn=10$^{-8}$ N and 1 $\text{\AA}$=10$^{-10}$ m.}
\begin{center}
\begin{tabular}{@{\hspace{8pt}} c @{\hspace{8pt}} | @{\hspace{6pt}} c @{\hspace{6pt}}  @{\hspace{6pt}} c   @{\hspace{6pt}}c  ||@{\hspace{6pt}} c  @{\hspace{6pt}}  | @{\hspace{6pt}} c   @{\hspace{6pt}}c   @{\hspace{6pt}} c @{\hspace{6pt}} @{\hspace{6pt}} c   @{\hspace{6pt}}c  | @{\hspace{6pt}} c @{\hspace{6pt}} | @{\hspace{6pt}} c @{\hspace{6pt}}}

\hline
Internal& {This work} & {H$\&$B}& & Normal& This work&&{H$\&$B}\\
\hline
$f_{rr}$ &4.1623(859)&	4.1736& &	$\omega_1$	&929.57&&935.2\\
$f_{rr'}$&1.8138(1022)&	1.9128	& &$\omega_3$	&635.32&&623.3\\
$f_{r\theta}$&0.215(177)&	0.3746	& &$\phi_{111}$	&-274.65&&-322.5\\
$f_{\theta\theta}$&1.1203(144)	& 1.1616&	&$\phi_{133}$&-283.85&&-305.8\\\
$f_{rrr}$&-32.698(1281)&-36.261& &$\phi_{3333}$&98.44&&122.4\\
$f_{rrr'}$&-4.988(928)&-7.015	& &$\phi_{1133}$&59.28&&72.0\\
$f_{rr\theta}$&-2.784(1332)&-3.340& &$\phi_{1111}$&27.54&&32.2\\\
$f_{rr'\theta}$&-0.301(662)&-0.918&&$\omega_2$&377.17&&384.9\\
$f_{r\theta\theta}$ &-3.467(1008)&-3.593&&$\phi_{122}$&-33.11&&-29.9\\
$f_{\theta\theta\theta}$ &-3.306(466)&-3.755&&$\phi_{112}$&-29.31&&-38.5\\
$f_{rrrr}$ &138.26(745)&165.66& &$\phi_{222}$&-78.86&&-79.6\\
 &&& &$\phi_{233}$&-49.80&&-44.4\\
\hline
$b$/cm$^{-1}$ &102.86(439) &	117.5\\
$\rho$ &0.3485(197)&0.4\\
$\theta_e$/deg.&103.80(3)&103.75\\
$r_e$/$\text{\AA}$  &1.5557(3)&1.5525\\
\hline

\end{tabular}
\end{center}
\label{tab:2D}
\end{table}

Before we present the results from a complete 3D fit using Eq.~(\ref{eq:Hamiltonian}), we first discuss the result obtained from a 2D fit model (excluding $\nu_2$ bending) originally developed by Hoy and Brand for the $\tilde{\text{C}}$ state of SO$_2$~\cite{Brand1978}. We follow their treatment, but we also include isotopologue information and rotational constants, as described in Section \ref{sec:methods}. The reason for the success of a reduced-dimension 2D fit model is that, although the $\nu_1$ and $\nu_3$ modes interact strongly, the $\nu_2$ mode remains approximately isolated below 2000 cm$^{-1}$~\cite{{Brand1978},{Yamanouchi1995},{Brand1976}}. The fit result is presented in Table~\ref{tab:2D}, along with the Hoy and Brand result~\cite{Brand1978}. The force field (from a 2D fit) derived in this work is qualitatively similar to the one obtained by Hoy and Brand~\cite{Brand1978}. Also listed are normal-mode force constants of S$^{16}$O$_2$ derived from the internal coordinate force field. The measured and calculated vibrational term values are included in Table~\ref{tab:2Denergy}. The quantum numbers and the subscript, $r$ or $l$, used in our vibrational assignments in Table~\ref{tab:2Denergy}, are related to the nodal patterns and the general shapes of the vibrational wavefunctions, respectively, which are explained in detail in Section~\ref{subsec:Fermi}.

\begin{table}[t]
\caption{Experimental (Obs.) and calculated (Cal.) vibrational term energies of states included in the 2D fit. For both isotopologues, the energy of the zero-point level has been subtracted from each term value. The observed~\cite{Johann1977} and calculated isotope shift between the zero-point vibrational levels of the $\tilde{\text{C}}$ state of S$^{16}$O$_2$ and S$^{18}$O$_2$ are -26.3 cm$^{-1}$ and -28.0 cm$^{-1}$, respectively. The asterisks indicate a mixing of wavefunctions between those of the nearest neighbor energy levels. The notations used in the vibrational assignments are explained in Section~\ref{subsec:Fermi}. Units in cm$^{-1}$.}   
\begin{center}
\begin{tabular}{@{\hspace{4pt}} c @{\hspace{4pt}} | @{\hspace{6pt}} c @{\hspace{6pt}} | @{\hspace{6pt}} c   @{\hspace{6pt}}c |@{\hspace{6pt}} c  @{\hspace{6pt}}  || @{\hspace{4pt}} c   @{\hspace{4pt}}c  | @{\hspace{6pt}} c @{\hspace{6pt}} | @{\hspace{6pt}} c   @{\hspace{6pt}}c  | @{\hspace{6pt}} c @{\hspace{6pt}}  @{\hspace{6pt}} c @{\hspace{6pt}}}

\hline
& Assig.& Obs. & & Cal.&&&Assig.& Obs. &&Cal. \\ 
\hline

S$^{16}$O$_2$	&	(0,0,0)$_r$	&	0	&&	0&S$^{16}$O$_2$&&(0,0,1)$_r$& 212.6 &&212.3\\
$a_1$ sym.	&	(0,0,2)$_r$	&	561.2	&&	560.7&$b_2$ sym.&&(0,0,3)$_r$& 890.9 &&891.1\\
	&	(1,0,0)$_l$	&	960.0	&&	960.5&&&(1,0,1)$_l$& 1261.4 &&1262.2\\
	&	(0,0,4)$_r$	&	1245.4	&&	1247.4&&&(0,0,5)$_r$& 1595.8&&1596.9\\
	&	(1,0,2)$_r$	&	1653.7	&&	1650.8&&&(1,0,3)$_r$& &&1996.7\\
	&	(2,0,0)$_l$*	&	1917.5	&&	1916.7&&&(0,0,7)$_r$*& &&2315.2\\
&	(0,0,6)$_r$*	&	1964.9	&&	1963.7&&&(2,0,1)$_l$*& &&2338.7\\
	&	(1,0,4)$_r$	&		&&	2371.7&&&(1,0,5)$_r$&  &&2729.4\\
	&	(0,0,8)$_r$*	&	2680.3	&&	2680.0&&& &  &&\\
	&	(2,0,2)$_r$*	&		&&	2727.7&&& & &&\\
	&	(3,0,0)$_l$	&	2920.6	&&	2921.9&&& & &&\\
\hline
S$^{18}$O$_2$	&	(0,0,2)$_r$	&	535.1		&&	534.7 &S$^{18}$O$_2$&&(0,0,1)$_r$& &&200.4\\
$a_1$ sym.	&	(1,0,0)$_l$	&	920.9$^a$	&&	921.1&$b_2$ sym.&&(0,0,3)$_r$&  &&852.2\\
	&	(1,0,2)$_r$	&1582.3$^a$	&&	1580.5&&&(1,0,1)$_l$& &&1206.4\\
	&	(2,0,0)$_l$*	&	1840.0$^a$	&&	1838.0 && & (0,0,5)$_r$&&&1531.6\\
&	(0,0,6)$_l$*	&	1880.2$^a$	&&	1884.8& && &  && \\
	&	(3,0,0)$_l$	& 2798.4$^a$	&&	2799.6  &&& & && \\
\hline
\multicolumn{6}{l}{a. Low resolution measurement~\cite{Brand1978}.}\\[-6pt]
\end{tabular}
\end{center}
\label{tab:2Denergy}
\end{table} 

\begin{figure}[b]
\includegraphics[width=5.5 in]{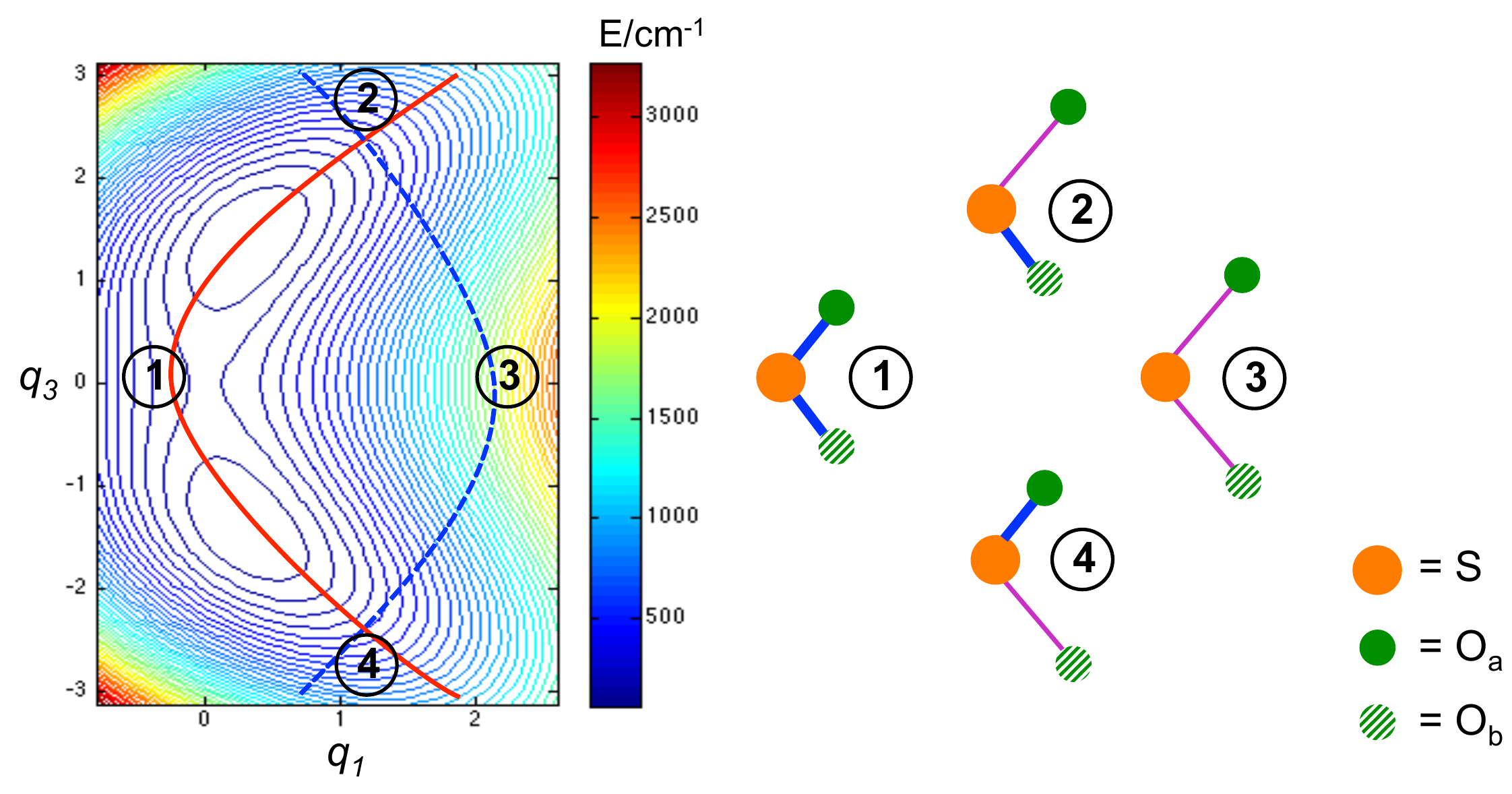}
\caption{Two-dimensional potential energy surface obtained from the 2D fit, along with the approximate geometries of the molecule at four different points on the PES. The two oxygen atoms are labeled and color-coded differently for clarity in the text. The red solid curve and the blue dashed curve on the PES are related to the nodal patterns of the wavefunctions, discussed in detail in Section~\ref{subsec:Fermi}.}
\label{fig:PES}
\end{figure}

Table~\ref{tab:2DGeometry} shows the $C_s$ equilibrium bond lengths obtained from our internal coordinate force field of the $\tilde{\text{C}}$ state of SO$_2$. The barrier on the PES is relatively low, but it is sufficient to produce a significant depression of the antisymmetric stretch fundamental frequency. The $\nu_3$ fundamental frequency, which is usually the highest among the three fundamental frequencies of symmetric triatomic molecules, is the lowest in the $\tilde{\text{C}}$ state of SO$_2$. The parameter, $\rho$, characterizes the curvatures at the two minima of the PES. The bottoms of the two wells on the PES would be nearly parabolic if $\rho=1.5$ (in the absence of cubic and higher-order anharmonicities)~\cite{Coon1966}. For $\rho<1.5$, which is true for the $\tilde{\text{C}}$ state of SO$_2$ (see Table~\ref{tab:2D}), starting from a minimum of the PES and moving along $q_3$, the potential should rise more steeply in the direction away from the barrier than in the direction towards the barrier. This is indeed the case for the $\tilde{\text{C}}$ state, as can be seen from the PES in Fig.~\ref{fig:PES}.

\begin{table}[t]
\caption{C$_s$ equilibrium geometry of the $\tilde{\text{C}}$ state of SO$_2$.}
\begin{center}
\begin{tabular}{@{\hspace{6pt}} c @{\hspace{6pt}} @{\hspace{6pt}} c @{\hspace{6pt}}  @{\hspace{6pt}} c   @{\hspace{6pt}}c  @{\hspace{6pt}} c  @{\hspace{6pt}}  @{\hspace{6pt}} c   @{\hspace{6pt}}c   @{\hspace{6pt}} c @{\hspace{6pt}}  @{\hspace{6pt}} c   @{\hspace{6pt}}c   @{\hspace{6pt}} c @{\hspace{6pt}} @{\hspace{6pt}} c @{\hspace{6pt}}}

\hline
 & This work & H$\&$B~\cite{Brand1978}   \\
\hline
$r$$_1$/$\text{\AA}$ & 1.642 & 1.639\\
$r$$_2$/$\text{\AA}$ & 1.494 & 1.491\\

\hline
\end{tabular}
\end{center}
\label{tab:2DGeometry}
\end{table}

\subsection{Fermi Resonance in the $\tilde{\text{C}}$ state of SO$_2$ and Vibrational Assignments}
\label{subsec:Fermi}

Due to the presence of a Gaussian hump along the $q_3$ coordinate, the antisymmetric stretch is grossly anharmonic and is poorly described by the harmonic basis set used to construct the vibrational Hamiltonian. Therefore, we use a set of $anharmonic$ basis states, $\ket{v_1,v_2,v_3}_\text{a}$, obtained from a first-order perturbation theory treatment of the vibrational Hamiltonian, where all terms other than the harmonic oscillator terms are treated as perturbations. Below 1000 cm$^{-1}$, the energies of the anharmonic states, i.e. the diagonal matrix elements of the vibrational Hamiltonian, are much closer to the observed eigenenergies than the energies of the harmonic basis states. For example, the energy of the $anharmonic$ $\ket{0,0,1}_\text{ah}$ basis state (317 cm$^{\text{-}1}$) is much closer than that of the $harmonic$ basis state $\ket{0,0,1}$ (635 cm$^{\text{-}1}$) to the observed $\nu_3$ eigenenergy (212 cm$^{\text{-}1}$). 

Due to the depression of the $\nu_3$ frequency, the $\ket{0,0,2}_\text{a}$ energy is close to that of the $\ket{1,0,0}_\text{a}$ state, with an energy separation of 200 cm$^{\text{-}1}$. Given the large $\phi_{133}$ constant ($-284$ cm$^{\text{-}1}$), anti-symmetric and symmetric stretching modes are mixed via Fermi resonance. This strong mixing was first noted by Hoy and Brand~\cite{{Brand1976},{Brand1978}}, and recognized by others in more recent studies~\cite{{Yamanouchi1995},{Sako1998},{Xie1999},{Johann1977},{Bludsky2000}}. A large $\phi_{133}$ force constant is not unusual in symmetric triatomic molecules. Consider, for example, values for the ground electronic state of H$_2$O ($-1785$ cm$^{-1}$)~\cite{Mills1974}, or SO$_2$ ($-319$ cm$^{-1}$)~\cite{Morino1964}. However, strong Fermi interaction between symmetric and antisymmetric triatomic stretching modes is unusual, because the harmonic stretching frequencies are not typically in 1:2 resonance. 

\begin{figure}
\includegraphics[width=5.8 in]{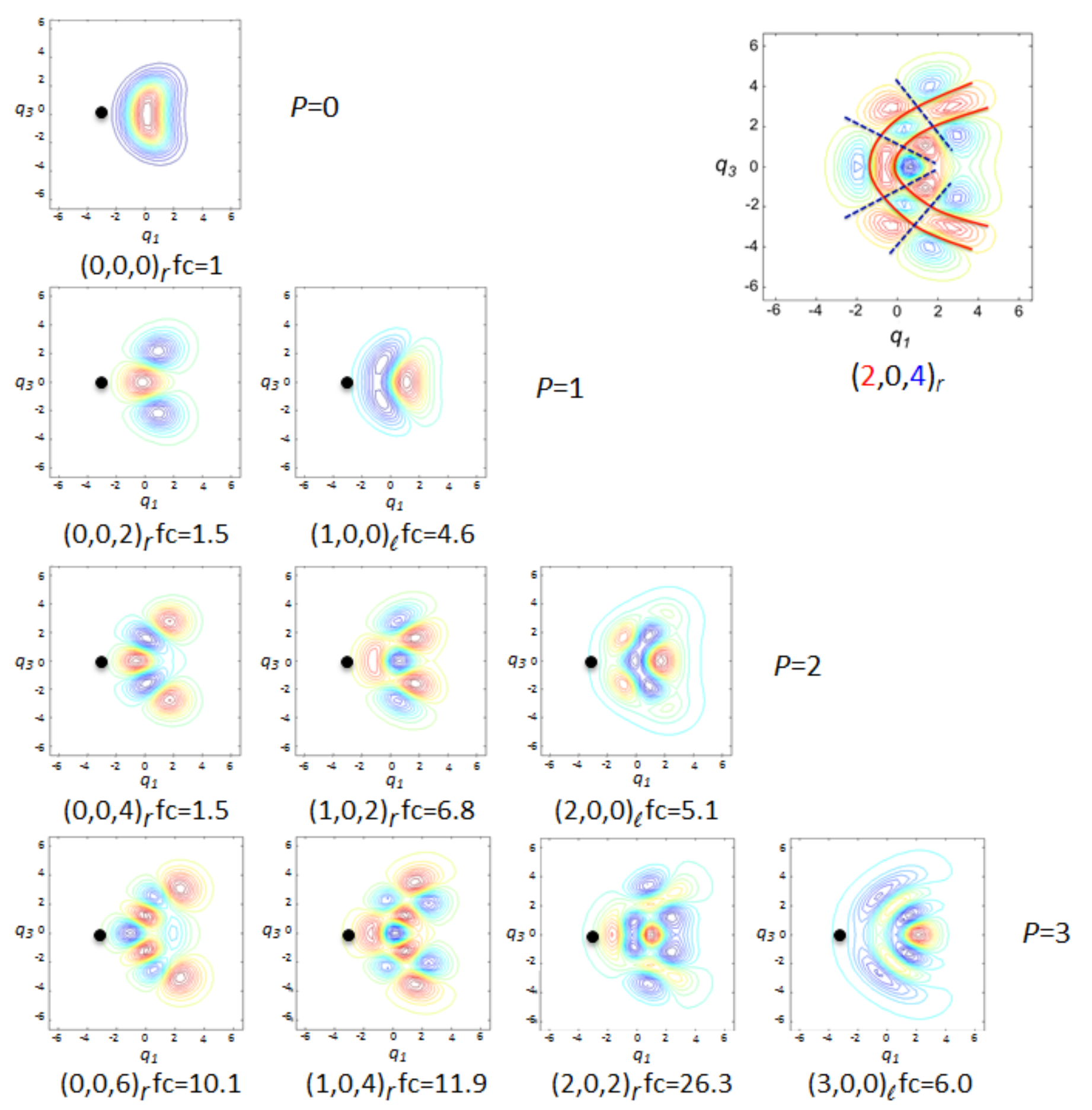}
\caption{$a_1$ vibrational symmetry wavefunctions of S$^{16}$O$_2$ from the 2D fit.  The semiclassical assignment (in parentheses) is given below each wavefunction, along with the calculated relative Franck-Condon factor (fc) for a transition from the zero-point level of the SO$_2$ $\tilde{\text{X}}$ state. The black dot on each figure is the approximate (center) location of the zero-point wavefunction of the S$^{16}$O$_2$ $\tilde{\text{X}}$ state. The vibrational wavefunction of the (2,0,4)$_r$ state in the $P$=4 polyad is plotted separately to illustrate how nodes in the eigenstates determine the $(v_\alpha,0,v_\beta)_r$ assignments. Within a given polyad (designated by polyad number, $P=v_\alpha+\frac{1}{2}v_\beta$), the energy increases for sub-figures from left to right.}
\label{fig:2Dpolyad}
\end{figure}

The effects of Fermi resonance on the semiclassical dynamics of molecules have been studied by Kellman and coworkers~\cite{{Kellman1986},{Kellman1988},{Kellman1990cat},{Kellman1990new},{Kellman1990}}. The standard procedure of labeling vibrational levels by normal-mode quantum numbers is inadequate and misleading for Fermi resonance systems. Kellman and coworkers provide an alternative assignment scheme based on the semiclassical dynamics~\cite{{Kellman1990cat},{Kellman1990new}}. An especially important feature of Kellman's assignment scheme is that one can make semiclassical vibrational assignments based on the nodal patterns of the wavefunction. This is particularly useful in the $\tilde{\text{C}}$ state of SO$_2$, because clear nodal patterns persist in many of the wavefunctions (available from Discrete Variable Representation calculations (DVR)), despite the fact that strong anharmonic effects prevent assignment of a dominant harmonic basis state, even at low vibrational energy. In Fig.~\ref{fig:2Dpolyad}, some of the $a_1$ vibrational symmetry wavefunctions of S$^{16}$O$_2$ obtained from our 2D fit are plotted, with assignments from Kellman's scheme discussed below. Additional discussion on the semiclassical dynamics of the $\tilde{\text{C}}$ state of SO$_2$ can be found in Appendix~\ref{app:kellman}. 

The shapes of the wavefunctions in Fig.~\ref{fig:2Dpolyad} are distorted from the shapes of normal-mode wavefunctions. Some of the wavefunction shapes are curved so that the nodal patterns extend along the red solid curve shown in Fig.~\ref{fig:PES}, while others extend along the blue dashed curve, perpendicular to the red curve. Wavefunctions with nodes organized along the red curve are given an `$r$' subscript in their assignment (indicating the wavefunctions ``open to the right''), e.g. $(0,0,6)_r$, while those with nodes organized along the blue curve are given an `$l$' subscript (indicating the wavefunctions ``open to the left''), e.g. $(3,0,0)_l$.

The three quantum numbers used in our assignment scheme, $(v_\alpha,v_2,v_\beta)$, are related to the nodal pattern of the wavefunctions. The second number, $v_2$, gives the number of bending quanta, which are uncoupled from stretching quanta in our 2D model. All vibrational levels displayed in this section and used in the 2D fit have $v_2=0$. The first and last quantum numbers describe the nodal pattern of the wavefunction in the 2D stretching plane. Using the wavefunction of the assigned (2,0,4)$_r$ state in Fig.~\ref{fig:2Dpolyad} as an example, the first number, $v_\alpha$, refers to the number of parallel red solid nodal curves that one can draw on the wavefunction, and the third number, $v_\beta$, is the number of blue dashed nodal lines, that cross the red solid nodal curves. The $\nu_{\alpha}$ mode, with $a_1$ vibrational symmetry, correlates to the symmetric stretching mode, $\nu_1$,  in the normal mode basis, and the $\nu_{\beta}$ mode, with $b_2$ vibrational symmetry, correlates to the anti-symmetric stretching mode, $\nu_3$.

Using our assignment scheme, the wavefunctions can be grouped according to the polyad number, $P=v_\alpha+\frac{1}{2}v_\beta$. There are $(P+1)$ levels that belong to a given polyad with polyad number $P$. A polyad consists of a group of systematically near-degenerate interacting zero-order states. For example, the three levels with polyad number $P=2$ result predominantly from three strongly anharmonically interacting zero-order wavefunctions $\ket{0,0,4}_\text{a}$, $\ket{1,0,2}_\text{a}$, and $\ket{2,0,0}_\text{a}$. 

\begin{figure}[t]
\includegraphics[width=3.2 in]{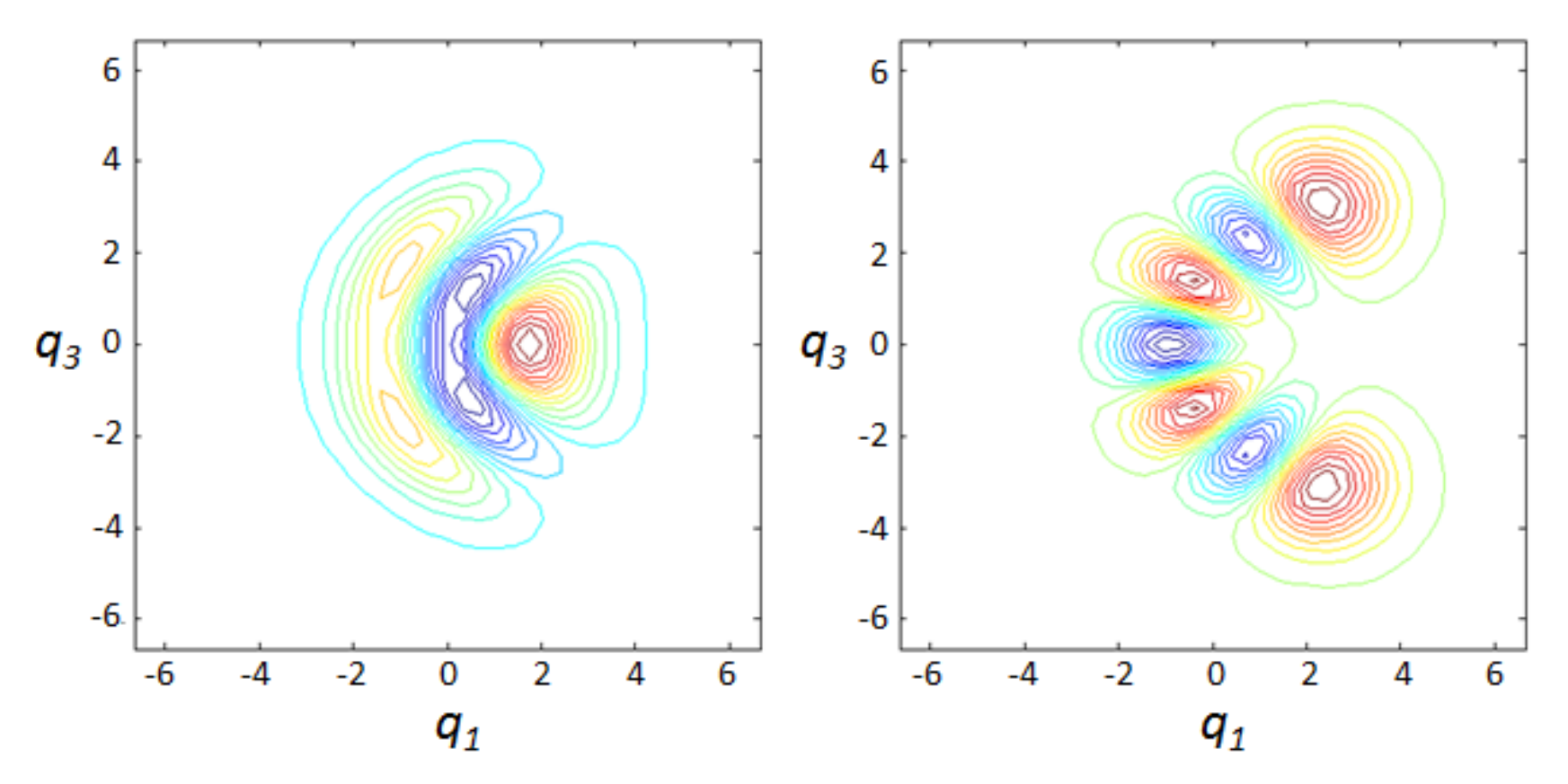}
\caption{Linear combinations of the (2,0,0)$_l$ and (0,0,6)$_r$ eigenstate wavefunctions (as in Fig.~\ref{fig:2Dpolyad}). The wavefunction on the left results from $0.93\,(2,0,0)_l-0.37\,(0,0,6)_r$ and the one on the right comes from $0.37\,(2,0,0)_l+0.93\,(0,0,6)_r$.}
\label{fig:lc_2D}
\end{figure}

The polyad number $P$ is not strictly conserved in the $\tilde{\text{C}}$ state of SO$_2$. Note that the (2,0,0)$_l$ and (0,0,6)$_r$ wavefunctions in Fig.~\ref{fig:2Dpolyad} appear to deviate from the expected shape of Kellman's Fermi resonance wavefunctions (e.g. there should not be a local maximum at $q_1=0, q_3=0$ for the (2,0,0)$_l$ wavefunction). However, if one takes a specific linear combination of (2,0,0)$_l$ and (0,0,6)$_r$ wavefunctions (see Fig.~\ref{fig:lc_2D}), the zero-order wavefunctions are restored, which indicates an interaction between the zero-order basis states. The $\phi_{1133}$ term, which has a magnitude of 60 cm$^{-1}$, is primarily responsible for the interaction. This Darling-Dennison interaction breaks the strict conservation of the polyad number, $P$. Similar interaction occurs between the zero-order (2,0,2)$_r$ and (0,0,8)$_r$ states (not shown in Fig.~\ref{fig:2Dpolyad}). The inter-polyad interaction that we see here is $not$ an artifact of the 2D nature of the fit, since it is observed in the wavefunctions obtained from the 3D fit as well (Section~\ref{subsec:vib_assign}). 

The peculiar shapes of the wavefunctions in Fig.~\ref{fig:2Dpolyad} have their origins in the shape of the PES. The PES has a kidney-bean shape (Fig.~\ref{fig:PES}), as a result of the large Fermi $\phi_{133}$ term. In the absence of strong Fermi-133 interaction, a double-well structure in the $q_3$ direction will cause a staggered energy pattern in the $normal$-$mode$ $(0,0,v_3)$ progression. The Fermi-133 resonance mixes the normal-modes to create Kellman-type modes, but the nodal coordinate along which the $(0,0,v_\beta)_r$ Kellman-type progression is organized--the red solid curve in Fig.~\ref{fig:PES}--also passes through both minima of the PES. As a result, levels in the $(0,0,v_\beta)_r$ progression, which extends along the new nodal coordinate, also exhibit staggering from the double-well minimum. The staggered energy pattern is obvious in Fig. 1 of the third paper of this series~\cite{Park2015vibronic}.

\subsection{Effects of Fermi Resonance on the Dynamics of $\tilde{\text{C}}$ state of SO$_2$}
\label{subsec:Fermi_effect}

The $r$- and $l$-type wavefunctions encode two different types of classical motions. As illustrated in Fig.~\ref{fig:PES}, molecules with the $r$-type motion pass through the $C_{2v}$ geometry with a shorter S-O bond length (see configuration $\textcircled{\small{1}}$) than those with the $l$-type motion (see $\textcircled{\small{3}}$). In the pure $r$-type motion, starting from $\textcircled{\small{2}}$, where the S-O$_a$ bond is stretched while the other bond is at approximately the bond length of the C$_{2v}$ configuration $\textcircled{\small{1}}$ ($R_1$), S-O$_a$ contracts to $R_1$. Then, while the S-O$_a$ bond remains at the bond length of $\textcircled{\small{1}}$, the S-O$_b$ bond elongates until the molecule reaches configuration $\textcircled{\small{4}}$. S-O$_a$ continues to stay at $R_1$, and S-O$_b$ contracts back to $R_1$. The molecule then returns to $\textcircled{\small{2}}$ by locally stretching the S-O$_a$ bond. The motion repeats itself. In the pure $l$-type motion, starting from $\textcircled{\small{2}}$, where the S-O$_b$ bond is contracted while the other bond is at approximately the bond-length of C$_{2v}$ configuration $\textcircled{\small{3}}$ ($R_3$), S-O$_b$ stretches to $R_3$. Then, while the S-O$_b$ bond remains at $R_3$, the S-O$_a$ bond contracts until the molecule reaches configuration $\textcircled{\small{4}}$. S-O$_b$ remains at $R_3$, while S-O$_a$ contracts back to $R_3$. The molecule then returns to $\textcircled{\small{2}}$ by locally contracting the S-O$_b$ bond. The motion then repeats itself. We emphasize that the semiclassical motion that results from the Fermi-133 resonance in the $\tilde{\text{C}}$ state is similar to but qualitatively different from the local stretching motion caused by the Darling-Dennison resonance at high excitation, e.g. in water. 

Sako et al.~\cite{Sako1998} first noted this behavior of the wavefunctions for the $\tilde{\text{C}}$ state of SO$_2$, although they did not give an explicit interpretation of the semiclassical motions encoded in the wavefunctions. In the same work~\cite{Sako1998}, it was noted that above the predissociation threshold, $r$-type levels dissociate more rapidly than $l$-type levels. They argue that the $r$-type wavefunctions, with intensity along the S-O dissociation coordinate, have better overlap with the dissociation continuum of the ground electronic surface, while the $l$-type wavefunctions have less overlap with the ground state continuum. The level-dependence of the predissociation rates of the $\tilde{\text{C}}$ state vibrational levels can also be understood in light of the semiclassical motions encoded in the wavefunctions. In the $r$-type motion, the stretching momentum is always localized in the S-O bond that is instantaneously longer, whereas the momentum is localized in the instantaneously shorter S-O bond in the $l$-type motion. As a result, vibrational levels with $r$-type semiclassical motion couple better to the reaction coordinate, while those with `$l$'-type motion waste their energy by visiting configuration $\textcircled{\small{3}}$ in Fig.~\ref{fig:PES}.

\subsection{Franck-Condon Factors of the SO$_2$ $\tilde{\text{C}}$-$\tilde{\text{X}}$ transition}

Also given in Fig.~\ref{fig:2Dpolyad} are calculated relative Franck-Condon factors (fc) from the zero-point level of the $\tilde{\text{X}}$ state of SO$_2$. The ground electronic state of SO$_2$ has a shorter equilibrium S-O bond length than the $\tilde{\text{C}}$ state. As a result, the symmetric stretching mode is Franck-Condon active. Ordinarily, one would not expect Franck-Condon activity in the antisymmetric stretching mode of a symmetric triatomic molecule. However, as shown in Fig.~\ref{fig:2Dpolyad}, for polyads with $P>1$, the highest energy member of the polyad, which correlates with a pure symmetric stretching zero-order state, does not have the largest Franck-Condon factor. The Franck-Condon intensities migrate toward lower polyad members as the polyad number increases, in agreement with the trend observed experimentally~\cite{Yamanouchi1995}. This can be understood in terms of the effect of the Fermi-133 resonance on the shapes of the wavefunctions. The black dot on each subfigure in Fig.~\ref{fig:2Dpolyad} indicates the approximate center location (at $q_1=-3.2$) of the $\tilde{\text{X}}$ state zero-point level wavefunction in the $q_1$-$q_3$ plane. Below 3000 cm$^{-1}$, the highest energy states of a given polyad all have $l$-type wavefunctions, which open up toward the negative $q_1$ direction and have almost no intensity at the Franck-Condon point, while lower members of a given polyad are $r$-type, which are curved toward the Franck-Condon active area. As a result, the highest member loses its Franck-Condon activity, while lower members gain significant intensity. 

\section{Three-dimensional fit}
\label{sec:3D fit}

\begin{table}
\caption{Internal force constants of the $\tilde{\text{C}}$ state of SO$_2$ obtained from a three-dimensional fit, along with normal-mode force constants of S$^{16}$O$_2$. Normal-mode force constants derived by Yamanouchi~\cite{Sako1998} are included for comparison. All of the normal-mode force constants have units of cm$^{-1}$. Internal force constants are given in the mdyn-$\text{\AA}$ unit system.}
\begin{center}
\begin{tabular}{@{\hspace{8pt}} c @{\hspace{8pt}} | @{\hspace{3pt}} c @{\hspace{3pt}} || @{\hspace{6pt}} c   @{\hspace{6pt}}c  |@{\hspace{6pt}} c  @{\hspace{6pt}}  @{\hspace{6pt}} c   @{\hspace{6pt}}c   @{\hspace{6pt}} c @{\hspace{6pt}} @{\hspace{6pt}} c   @{\hspace{6pt}}c   @{\hspace{6pt}} c @{\hspace{6pt}} | @{\hspace{6pt}} c @{\hspace{6pt}}}

\hline
Internal& {This work} & Normal& & This work& Yamanouchi~\cite{Sako1998}\\
\hline
$f_{rr}$ &3.9326(353)&	$\omega_1$& & 938.03 &942.6\\
$f_{rr'}$&2.0185(476)&	$\omega_2$	& &392.28	&389.13\\
$f_{r\theta}$&0.093(58)&	$\omega_3$	& &573.56	&589.6\\
$f_{\theta\theta}$&1.2250(97)	& $\phi_{111}$&	&-283.72&-306.0\\
$f_{rrr}$&-32.080(1044)&$\phi_{133}$& &-300.21&-288.6\\
$f_{rrr'}$&-5.763(924)&$\phi_{112}$	& &-46.143&-22.16\\
$f_{rr\theta}$&-1.702(432)&$\phi_{122}$& &0.381&16.26\\
$f_{rr'\theta}$&-0.354(176)&$\phi_{222}$&&-85.375&-65.538\\
$f_{r\theta\theta}$ &-2.156(217)&$\phi_{233}$&&-48.584&-43.48\\
$f_{\theta\theta\theta}$ &-3.521(64)&$\phi_{1111}$&&57.321&33.12\\
$f_{rrrr}$ &149.89(1049)&$\phi_{1133}$& &52.237&53.52\\
$f_{rrrr'}$ &-4.226(8783)&$\phi_{3333}$& &223.05&177.36\\
$f_{rrr'r'}$ &29.60(1034)&$\phi_{1122}$& &-0.458&-21.52\\
$f_{rr\theta\theta}$ &12.21(226)&$\phi_{2222}$& &8.276&\\
$f_{rr'\theta\theta}$ &7.26(210)&$\phi_{2233}$& &-8.827&\\
$f_{\theta\theta\theta\theta}$ &9.88(169)&$\phi_{1222}$& &8.110&\\
$f_{rrr\theta}$ &5.78(379)&$\phi_{1112}$& &0.086&\\
$f_{rrr'\theta}$ &-5.15(478)&$\phi_{1233}$& &10.294&\\
$f_{r\theta\theta\theta}$ &3.90(272)& && &\\
\hline
$b$/cm$^{-1}$ &\multicolumn{4}{l}{\,90.39(180)}&117.5\footnote{Constrained to the 2D fit value of Ref ~\onlinecite{Brand1978}.}\\
$\rho$ &\multicolumn{4}{l}{\,\,\,\,\,\,\,\,0.35\footnote{Constrained.}}&0.4$^{a}$\\\
$\theta_e$/deg.&\multicolumn{4}{l}{\,\,\,103.80(2)}& 103.75$^{a}$\\
$r_e$/$\text{\AA}$  &\multicolumn{4}{l}{\,\,\,1.5557(3)}&1.5525$^{a}$\\
\hline

\end{tabular}
\end{center}
\label{tab:3D}
\end{table}

The internal coordinate force field obtained from our 3D fit is presented in Table~{\ref{tab:3D}, along with the normal-mode force constants derived for S$^{16}$O$_2$, for comparison with values derived by Yamanouchi~\cite{Sako1998}. In our 3D fit, despite inclusion of rotational and isotopologue information, all three parameters that characterize the Gaussian hump in Eq.~(\ref{eq:C1}) and (\ref{eq:C2}), $b$, $\omega_3$ (which in our internal coordinate force field is determined by $f_{rr}$ and $f_{rr'}$), and $\rho$ are strongly correlated ($>$0.95 correlation parameters among them). To break this correlation, additional $b_2$ vibrational symmetry levels, especially the (0,0,1)$_r$ levels of different isotopologues, must be measured and included in the fit. In the absence of those isotopologue data, we have fixed the value of $\rho$ to 0.35, which is the value we obtained from our two-dimensional fit. The value of $\rho$ was better determined in the 2D fit, due to constraints made to obtain the force field (although we cannot guarantee the accuracy of $\rho$ obtained from the 2D fit). By constraining the value of $\rho$, the correlation among $B$, $f_{rr}$, and $f_{rr'}$ is much reduced ($<$0.5). We must emphasize that the uncertainties of the fit parameters listed in Table~{\ref{tab:3D} are only statistical uncertainties of the fit, which do not take into account correlation effects. The actual uncertainties in some of the parameters might realistically be 5 to 10 times larger~\cite{Mills1974}. Information from other isotopologues would certainly reduce correlation. Alternatively, high-level quantum calculations might provide better constraints on some of the fit parameters, which would allow us to obtain a more physical and accurate internal coordinate force field fit. In Tables~\ref{tab:3DEnergy16}--\ref{tab:Cor_matrix}, the measured and calculated values of the observables from our 3D calculation are listed. 

For all of the vibrational term values included in the fit, the difference between experiment and fit is less than 1.9 cm$^{-1}$ and the rms error is 0.9 cm$^{-1}$. The observed level at 2224.9 cm$^{-1}$~\cite{Yamanouchi1995} is not included in our fit. Based on the energy, the only possible assignment is (0,6,0)$_r$, which, according to our derived force field, is predicted at 2208 cm$^{-1}$ (see Table~\ref{tab:3DEnergy16}). However, the 2224.9 cm$^{-1}$ level seems unlikely to correspond to the (0,6,0)$_r$ state. The $A$ rotational constant of (0,6,0)$_r$ is predicted to be around 1.27 cm$^{-1}$, given the large positive $\alpha_2^a$ constant (0.02 cm$^{-1}$). The experimentally derived $A$ rotational constant of the 2224.9 cm$^{-1}$ level is, however, only 1.1226(159) cm$^{-1}$~\cite{Yamanouchi1995}, which is significantly smaller than $any$ of the $A$ constants of the $\tilde{\text{C}}$ state of SO$_2$. In addition, the observed and calculated trend in the Franck-Condon factors suggests that the (0,6,0)$_r$ level should be too weak to be observed in the spectrum. Using our force field, it is also clear that the 2224.9 cm$^{-1}$ level is not due to a hot-band transition; nor can it be a level from $^{34}$SO$_2$, which has about 5$\%$ natural abundance. We believe that the 2224.9 cm$^{-1}$ level may be an interloper from another electronic state that borrows transition intensity from the $\tilde{\text{C}}$-state. Further characterizations of this level (e.g. fluorescence lifetime and magnetic field response measurements) are necessary to confirm its identity. Considering all the uncertainties pertaining to this level, we have excluded it from our fit model.

\begin{table}
\caption{Experimental (Obs.) and calculated (Cal.) vibrational term energies of $a_1$ symmetry states of S$^{16}$O$_2$, along with their vibrational assignments. The energy of the $\tilde{\text{C}}$-state origin has been subtracted from each term value. Unless otherwise stated, the experimental term values are from Ref ~\onlinecite{Yamanouchi1995}. Units in cm$^{-1}$. Levels are labeled according to the degree of perturbation in the Kellman basis (explained in Section~\ref{subsec:vib_assign}). Levels with a single Kellman basis state coefficient greater than 0.9 are considered minimally perturbed (no label); `$\ast$' indicates a coefficient of 0.8--0.9; `$\dagger$' indicates a coefficient of 0.7--0.8; `$\ddagger$' indicates coefficients $<$0.7 (no nominal assignment possible). Superscript numbers after the assignments indicate eigenstates that result from the same set of interacting Kellman basis states, e.g. (0,2,4)$_r$ and (0,0,6)$_r$ interact to yield the eigenstates at 1958 cm$^{-1}$ and 1965 cm$^{-1}$ (both with superscript 1).}    
\begin{center}
\begin{tabular}{@{\hspace{7pt}} c@{\hspace{7pt}} c@{\hspace{7pt}} c@{\hspace{7pt}} |@{\hspace{7pt}} c@{\hspace{7pt}} c@{\hspace{7pt}} c@{\hspace{7pt}}|@{\hspace{7pt}} c@{\hspace{7pt}} c@{\hspace{7pt}} c@{\hspace{7pt}} }

\hline 
\multicolumn{1}{@{\hspace{7pt}} c@{\hspace{7pt}} }{\textbf{Assig.}} & \multicolumn{1}{c@{\hspace{7pt}} }{\textbf{Obs.}} & \multicolumn{1}{c@{\hspace{7pt}} |@{\hspace{7pt}} }{\textbf{Cal.}} & \multicolumn{1}{c@{\hspace{7pt}} }{\textbf{Assig.}} & \multicolumn{1}{c@{\hspace{7pt}} }{\textbf{Obs.}} & \multicolumn{1}{c@{\hspace{7pt}} |@{\hspace{7pt}}}{\textbf{Cal.}} & \multicolumn{1}{c@{\hspace{7pt}} }{\textbf{Assig.}} & \multicolumn{1}{c@{\hspace{7pt}} }{\textbf{Obs.}} & \multicolumn{1}{c@{\hspace{7pt}}}{\textbf{Cal.}} \\
\hline

 (0,0,0)$_r$ & 0 	& 0	   & (0,2,4)$_r*^1$ &  	& 1958.2  & (0,0,8)$_r$ &2680.3 & 2681.2\\
 (0,1,0)$_r$ & 377.5 & 377.3&(0,0,6)$_r*^1$ & 1964.9 & 1965.4 & $\ddagger^8$	& 		  & 2729.3 \\
 (0,0,2)$_r$ & 561.2 & 561.2& (1,1,2)$_r$ & 2018.9 & 2019.8 & (0,6,2)$_r\dagger^6$	&	 & 2739.5\\
 (0,2,0)$_r$ & 751.5 & 751.0& (0,4,2)$_r*^2$ &  		& 2024.8 & $\ddagger^8$	&2743.0$^{a,c}$  & 2744.7 \\
 (0,1,2)$_r$ & 932.0 & 930.9 & (1,3,0)$_l$ & 2084.3 & 2084.6  & $\ddagger^8$	& 2762.1 & 2763.6\\
 (1,0,0)$_l$ & 960.0 & 960.8 & (0,6,0)$_r*^3$ &  & 2207.3  & (1,5,0)$_l$ & 		& 2817.5 \\
 (0,3,0)$_r$ & 1122.0 & 1121.1 & (2,1,0)$_l$ & 2285.3 & 2284.4 & (0,8,0)$_r$* & 		& 2910.1\\
 (0,0,4)$_r$ & 1245.4 & 1246.6 &  (0,3,4)$_r\dagger^4$ & 2308.7 & 2310.0 & (3,0,0)$_l$ &2920.6 & 2921.5 \\
 (0,2,2)$_r$ & 1300.0 & 1298.1 & (0,1,6)$_r\dagger ^4$ & 2321.8 & 2320.8 & $\ddagger$	  & 3118.6$^{a}$ & 3119.8\\
 (1,1,0)$_l$ & 1337.9 & 1338.5 &(1,2,2)$_r*^5$ & 2370$^{a,b}$ & 2378.3 & (1,0,6)$_r\dagger$  & 3136.4$^{a}$ & 3141.0\\
 (0,4,0)$_r$ &  	& 1487.2 & (0,5,2)$_r*^3$ &  	     & 2383.8 & (3,1,0)$_l$ & 3281.8$^{a}$ & 3283.9\\
 (0,1,4)$_r$ & 1604.3 & 1604.9 &(1,0,4)$_r*^5$ & 2394.3 & 2393.8 & $\ddagger$     &3494.8$^{a}$ & 3500.4\\
 (1,0,2)$_r$ & 1653.7 & 1653.9 & (1,4,0)$_l$ & 2452.6 & 2452.8 & (2,0,4)$_r$ & 3526.0$^{a}$ & 3528.8\\
 (0,3,2)$_r$ &  	& 1662.8 & (0,7,0)$_r*^6$ &  		& 2560.9 & (3,2,0)$_l$ & 3640.5$^{a}$ & 3643.9\\
 (1,2,0)$_l$ & 1712.7 & 1713.1 & $\ddagger^7$	& 2644.3 & 2643.7 & (3,0,2)$_r$ &3763.9$^{a}$ &3762.3\\
 (0,5,0)$_r$$^2$ &  		& 1849.3 &  $\ddagger^7$	&2663.5  & 2662.8 & $\ddagger$    & 3887.7$^{a}$ & 3898.5\\
 (2,0,0)$_l$ & 1917.5 & 1916.3 &$\ddagger^7$	&  		& 2673.4 & (3,3,0)$_l$  & 3996.8$^{a}$ & 4001.2\\
\hline
\multicolumn{6}{l}{a. Not included in the fit.}\\[-6pt]
\multicolumn{6}{l}{b. Low resolution measurement~\cite{Ebata1988}.}\\[-6pt]
\multicolumn{6}{l}{c. MODR result~\cite{Park2015}.}\\[-6pt]

\end{tabular}
\end{center}
\label{tab:3DEnergy16}
\end{table}

\begin{table}[t]
\caption{Experimental (Obs.) and calculated (Cal.) vibrational term energies of $b_2$ symmetry levels of S$^{16}$O$_2$, as well as both $a_1$ and $b_2$ levels of S$^{18}$O$_2$. The energy of the zero-point level has been subtracted from each term value. The observed~\cite{Johann1977} and calculated isotope shift between the zero-point vibrational levels of the $\tilde{\text{C}}$ state of S$^{16}$O$_2$ and S$^{18}$O$_2$ are -26.3 cm$^{-1}$ and -27.7 cm$^{-1}$, respectively. See the caption of Table~\ref{tab:3DEnergy16} and Section~\ref{subsec:vib_assign} for details regarding the meaning of the notations in the assignments.}    
\begin{center}
\begin{tabular}{@{\hspace{4pt}} c @{\hspace{4pt}} | @{\hspace{6pt}} c @{\hspace{6pt}} | @{\hspace{6pt}} c   @{\hspace{6pt}}c |@{\hspace{6pt}} c  @{\hspace{6pt}}  || @{\hspace{4pt}} c   @{\hspace{4pt}}c  | @{\hspace{6pt}} c @{\hspace{6pt}} | @{\hspace{6pt}} c   @{\hspace{6pt}}c  | @{\hspace{6pt}} c @{\hspace{6pt}}  @{\hspace{6pt}} c @{\hspace{6pt}}}

\hline\hline
& Assg.& Obs. & & Cal.&&&Assig.& Obs. &&Cal.\\ 
\hline

S$^{16}$O$_2$	&	(0,0,1)$_r$	&	212.6	&&	212.8&S$^{18}$O$_2$&&(0,1,0)$_r$& 359.5 && 359.6\\
$b_2$ sym.	&	(0,1,1)$_r$	&	582.2	&&	583.2&$a_1$ sym.&&(0,0,2)$_r$& 535.1 && 534.5\\
	&	(0,0,3)$_r$	&	891.0	&&	891.1 &&&(0,2,0)$_r$& 715.2\footnote{Based on low resolution band-head measurement.} &&715.9\\
	&	(0,2,1)$_r$	&	949.1	&&	950.4&&&(1,0,0)$_l$& 920.9$^a$ &&922.3\\
	&	(0,1,3)$_r*^1$	&	1252.3	&&	1251.8&&&(1,0,2)$_r$& 1582.3$^a$\footnote{Not included in the fit.} &&1581.9\\
	&	(1,0,1)$_l*^1$	&	1261.3	&&	1259.7&&&(2,0,0)$_l$& 1840.0$^a$$^b$ &&1838.3\\
	&	(0,3,1)$_r$	&	1313.2	&&	1314.3&&&(0,0,6)$_r$& 1880.2$^a$$^b$&&1882.6\\
	&	(0,0,5)$_r$	&	1595.8	&&	1595.5&&& (3,0,0)$_l$& 2798.4$^a$$^b$ &&2798.7\\
	&	(0,2,3)$_r*^2$	&			&&	1612.0&$b_2$ sym.&&(0,0,1)$_r$&  &&200.6\\
	&	(1,1,1)$_l*^2$	&					&&	1628.4&&&(0,1,1)$_r$ & &&553.8\\
	&	(1,0,5)$_r$	&	2754.7	&&	2752.8&&& & &&\\

\hline\hline
\end{tabular}
\end{center}
\label{tab:3Denergy_rest}
\end{table}

For the rotational constants included in the fit (Table~\ref{tab:3Drot}), the calculated values fall within (or very close to) the 2$\sigma$ uncertainties of the experimentally derived values, except for the $A$ and $B$ constants of the (1,0,0)$_l$ level of S$^{16}$O$_2$. The (1,0,0)$_l$ level is Coriolis-coupled to the close-lying (0,2,1)$_r$. Due to lack of high-$J$ data points for (0,2,1)$_r$, which are crucial to deperturbation of the Coriolis interactions, the derived rotational constants of (1,0,0)$_l$~\cite{Park2015SO2}, especially the $B$ and $C$ constants, are likely not fully deperturbed. In addition, the total error (0.073 cm$^{-1}$) of the fit to the Coriolis-interacting tetrad containing (1,0,0)$_l$ and (0,2,1)$_r$ is significantly larger than the calibration error (0.02 cm$^{-1}$), and not all parameters are fit simultaneously. Therefore, the real uncertainties in the derived rotational constants of (1,0,0)$_l$ can be significantly larger than the the statistical uncertainties. Overall, the fit to the centrifugal distortion coefficients given in Table~\ref{tab:3Dcent} is quite good, although some of the centrifugal distortion coefficients included in the fit (e.g. $\Delta_K$ of both isotopologues) fall outside of the 3$\sigma$ uncertainties of the experimentally derived values~\cite{Johann1977}. The treatments of the centrifugal distortion coefficients used in this work follow the treatments developed by Hoy and Brand~\cite{Brand1978}, which take into account the effect of the double-well on the centrifugal distortion coeffcients, but neglect other smaller anharmonic effects. This is likely the source of discrepancies between the observed and calculated values of some of the centrifugal distortion coefficients. 

\LTcapwidth=\textwidth
\begin{center}
\begin{longtable}{@{\hspace{7pt}} c||@{\hspace{7pt}} c@{\hspace{7pt}} c@{\hspace{7pt}} @{\hspace{7pt}} c@{\hspace{7pt}} c@{\hspace{7pt}} c@{\hspace{7pt}}}
\caption{Experimental and calculated rotational constants of S$^{16}$O$_2$. The experimentally derived rotational constants are given below the calculated values. Numbers below the vibrational assignments are the vibrational energy of that level. Values in bold face are included in our fit. Units in cm$^{-1}$. $2\sigma$ uncertainties are given for experimentally derived rotational constants. The meanings of \textbf{$C_{dp}$}, $C_{Cor}$, and \textbf{$C_{p}$} are defined in Section~\ref{sec:methods}.} \label{tab:3Drot} \\

\hline \multicolumn{1}{@{\hspace{7pt}} c||@{\hspace{7pt}} }{{Assig.}} & \multicolumn{1}{c@{\hspace{7pt}} }{$A$} & \multicolumn{1}{c@{\hspace{7pt}} @{\hspace{7pt}} }{$B$} & \multicolumn{1}{c@{\hspace{7pt}} }{{$C_{dp}$}} & \multicolumn{1}{c@{\hspace{7pt}} }{{$C_{Cor}$}} & \multicolumn{1}{c@{\hspace{7pt}}}{{$C_{p}$}} \\ \hline 
\endfirsthead

\multicolumn{6}{c}%
{{\tablename\ \thetable -- \textit{continued from previous page}}} \\
\hline \multicolumn{1}{@{\hspace{7pt}} c||@{\hspace{7pt}} }{{Assig.}} & \multicolumn{1}{c@{\hspace{7pt}} }{$A$} & \multicolumn{1}{c@{\hspace{7pt}} @{\hspace{7pt}} }{$B$} & \multicolumn{1}{c@{\hspace{7pt}} }{{$C_{dp}$}} & \multicolumn{1}{c@{\hspace{7pt}} }{{$C_{Cor}$}} & \multicolumn{1}{c@{\hspace{7pt}}}{{$C_{p}$}} \\ \hline 
\endhead

\multicolumn{6}{r}{{\textit{Continued on next page}}} \\
\endfoot

\endlastfoot

 		(0,0,0)$_r$	&	1.1505	&	0.3475	&	0.2658	&	0.0000	&	0.2658	\\
		0		&	$\bf{1.1505(1)}$	&	$\bf{0.3475(1)}$	&			&		&	0.2654(1)	\\\hline
		(0,0,1)$_r$	&	1.1466	&	0.3447	&	0.2631	&	-0.0015	&	0.2616	\\
		213		&	$\bf{1.1474(16)}$	&	$\bf{0.3444(5)}$	&		&		&	0.2614(4)	\\\hline
		(0,1,0)$_r$	&	1.1705	&	0.3460	&	0.2650	&	0.0013	&	0.2663	\\
		378		&	$\bf{1.1705(1)}$	&	$\bf{0.3459(1)}$	&		&		&	0.2658(1)	\\\hline
		(0,0,2)$_r$	&	1.1447	&	0.3427	&	0.2614	&	-0.0168	&	0.2445	\\
		561		&	1.1443(4)	&	0.3429(1)	&	0.2615(1)	&		&	0.2457(1)	\\\hline
		(0,1,1)$_r$	&	1.1672	&	0.3432	&	0.2623	&	0.0136	&	0.2759	\\
		582		&	1.1695(11)	&	0.3382(2)	&	0.2596(4)	&		&	0.2743(7)	\\\hline
		(0,2,0)$_r$	&	1.1905	&	0.3443	&	0.2641	&	0.0025	&	0.2666	\\
		752		&	1.1914(1)	&	0.3443(1)	&			&		&	0.2657(1)	\\\hline
		(0,0,3)$_r$	&	1.1424	&	0.3403	&	0.2594	&	-0.0105	&	0.2488	\\
		891		&	1.1432(19)	&	0.3405(5)	&	0.2595(10)	&		&	0.2498(4)	\\\hline
		(0,1,2)$_r$	&	1.1659	&	0.3413	&	0.2607	&	-0.0290	&	0.2317	\\
		932		&	1.1627(3)	&	0.3359(3)	&	0.2574(5)	&		&	0.242(4)	\\\hline
		(0,2,1)$_r$	&	1.1877	&	0.3416	&	0.2615	&	0.0334	&	0.2949	\\
		949		&	1.1908(17)	&	0.3430(6)	&	0.2625(13)	&		&	0.2906(6)	\\\hline
		(1,0,0)$_l$	&	1.1485	&	0.3444	&	0.2637	&	0.0008	&	0.2645	\\
		961		&	$\bf{1.1480(2)}$	&	$\bf{0.3456(1)}$	&	0.2643(2)	&		&	0.266(16)	\\\hline
		(0,3,0)$_r$	&	1.211	&	0.3426	&	0.2632	&	0.0034	&	0.2666	\\
		1122		&	1.209(12)	&	0.3419(24)	&		&		&	0.2650(21)	\\\hline
		(0,0,4)$_r$	&	1.1412	&	0.3383	&	0.2578	&	-0.0856	&	0.1722	\\
		1245		&	1.1389(16)	&	0.3398(5)	&	0.2586(9)	&		&	0.2008(18)	\\\hline
		(0,1,3)$_r$	&	1.1581	&	0.3395	&	0.2592	&	0.0655	&	0.3246	\\
		1252		&	1.1670(30)	&	0.3404(6)	&	0.2601(11)	&		&	0.2926(20)\\\hline
		(1,0,1)$_l$	&	1.1488	&	0.3404	&	0.2597	&	-0.0031	&	0.2566	\\
		1261		&	1.1462(12)	&	0.3420(3)	&		&		&	0.2558(2)	\\\hline		
		(0,2,2)$_r$	&	1.187	&	0.3399	&	0.2601	&	-0.0530	&	0.2071	\\
		1300		&	1.1861(98)	&	0.3365(11)	&	0.2580(22)	&		&	0.2069(23)	\\\hline
		(0,3,1)$_r$	&	1.2084	&	0.3400	&	0.2607	&	0.0643	&	0.3250	\\
		1313		&	1.2140(25)	&	0.3390(9)	&	0.2604(17)	&		&	0.3188(14)\\\hline
		(1,1,0)$_l$	&	1.169	&	0.3428	&	0.2628	&	0.0047	&	0.2675	\\
		1338		&	1.182(26)	&	0.3433(24)	&		&		&	0.2685(10)	\\\hline
		(0,0,5)$_r$	&	1.1414	&	0.3361	&	0.2561	&	-0.0425	&	0.2137	\\
		1596		&	1.1399(27)	&	0.3384(8)	&	0.2574(17)	&		&	0.2128(10)	\\\hline
		(0,1,4)$_r$	&	1.163	&	0.3368	&	0.2571	&	-0.1147	&	0.1444	\\
		1604		&	1.161(26)	&	0.3400(25)	&	0.2588(50)	&		&	0.1795(11)\\\hline
		(0,2,3)$_r$	& 	1.1815		& 0.3377		&0.2582	 & 0.1269	  &0.3851\\
		1612	$^a$& 	1.1879$^b$ & 0.3388$^b$	&	0.2600$^b$ & 	  &\\
		\hline
\multicolumn{6}{l}{a. Not directly observed. Band-origin calculated from the fit.}\\
\multicolumn{6}{l}{b. Constrained in the rotational fit~\cite{Park2015SO2}.}\\

\end{longtable}
\end{center}

\newpage

\begin{table}[t]
\caption{Observed~\cite{Johann1977} and calculated rotational constants of S$^{18}$O$_2$. 2$\sigma$ uncertainties are given for the experimentally derived values.}
\begin{center}
\begin{tabular}{@{\hspace{6pt}} c @{\hspace{6pt}} |@{\hspace{6pt}} c @{\hspace{6pt}}  @{\hspace{6pt}} c   @{\hspace{6pt}} @{\hspace{6pt}} c  @{\hspace{6pt}} c  @{\hspace{6pt}}  @{\hspace{6pt}} c   @{\hspace{6pt}}c   @{\hspace{6pt}} c @{\hspace{6pt}}  @{\hspace{6pt}} c   @{\hspace{6pt}}c   @{\hspace{6pt}} c @{\hspace{6pt}} @{\hspace{6pt}} c @{\hspace{6pt}}}

\hline
Assig.&$A$&$B$&$C$\\ \hline
(0,0,0)$_r$ & 1.0862 & 0.3089 & 0.2396 \\
&1.0863(2)&0.3089(1)&0.2392(1)\footnote{Not included in the fit}\\
(0,1,0)$_r$ & 1.1039 & 0.3076 & 0.2399 \\
 & 1.1038(1) &0.3077(1) & 0.2391(1)$^a$\\
\hline

\end{tabular}
\end{center}
\label{tab:3Drot2}
\end{table}

\begin{table}[t]
\caption{Observed~\cite{Johann1977} and calculated quartic centrifugal distortion coefficients for the zero-point levels of S$^{16}$O$_2$ and S$^{18}$O$_2$, using Watson's A reduction in the I$^{\text{r}}$ representation. 3$\sigma$ uncertainties are given for the experimentally derived values. Units in cm$^{-1}$.}
\begin{center}
\begin{tabular}{@{\hspace{6pt}} c @{\hspace{6pt}} |@{\hspace{6pt}} c @{\hspace{6pt}}  @{\hspace{6pt}} c   @{\hspace{6pt}} |@{\hspace{6pt}} c  @{\hspace{6pt}} c  @{\hspace{6pt}}  @{\hspace{6pt}} c   @{\hspace{6pt}}c   @{\hspace{6pt}} c @{\hspace{6pt}}  @{\hspace{6pt}} c   @{\hspace{6pt}}c   @{\hspace{6pt}} c @{\hspace{6pt}} @{\hspace{6pt}} c @{\hspace{6pt}}}

\multicolumn{1}{c}{}&\multicolumn{2}{c}{S$^{16}$O$_2$}&\multicolumn{2}{c}{S$^{18}$O$_2$}\\
\hline
&Obs.&Cal.&Obs.&Cal.\\ \hline
$10^7\Delta_J$ & 4.98(77)&4.44&4.07(11)&3.55\\
$10^7\Delta_{JK}$ & 129.2(60)&136.0&113.7(23)&115.8\\
$10^7\Delta_K$ & 73.8(103)&55.8&75.0(125)&55.7\\
$10^7\delta_J$ & 1.60(52) &1.53&1.40(9)&1.18\\
$10^7\delta_K$ & 84.0(26) &92.8&78.4(30)&79.1\\
\hline

\end{tabular}
\end{center}
\label{tab:3Dcent}
\end{table}

\begin{table}[t]
\caption{Calculated and experimentally-determined~\cite{Park2015SO2} $c$-axis Coriolis matrix elements, $t_1$ (in cm$^{-1}$ units). The harmonic predictions (reproduced from Table X of Ref ~\onlinecite{Park2015SO2}) are also listed for comparison. Values in parentheses are the 2$\sigma$ uncertainty of the final significant digits.}
\begin{center}
\begin{tabular}{@{\hspace{6pt}} c @{\hspace{6pt}} @{\hspace{6pt}} c @{\hspace{6pt}}  @{\hspace{6pt}} c   @{\hspace{6pt}} @{\hspace{6pt}}c  @{\hspace{6pt}}@{\hspace{6pt}} c  @{\hspace{6pt}}  @{\hspace{6pt}} c   @{\hspace{6pt}}c   @{\hspace{6pt}} c @{\hspace{6pt}}  @{\hspace{6pt}} c   @{\hspace{6pt}}c   @{\hspace{6pt}} c @{\hspace{6pt}} @{\hspace{6pt}} c @{\hspace{6pt}}}

\hline
&Expt.&Cal.&Harmonic\\ \hline
$(0,1,1)_r$-$(0,0,2)_r$ & 0.2978(7)&0.3040&0.3819\\
$(0,0,3)_r$-$(0,1,2)_r$ & 0.3250(89)&0.3216&0.4677\\
$(0,2,1)_r$-$(0,1,2)_r$ & 0.3532(44)&0.4357&0.5401\\
$(0,1,3)_r$-$(0,0,4)_r$ & 0.3463(14)&0.3211&0.5401\\
$(0,1,3)_r$-$(0,2,2)_r$ & 0.4528(99)&0.3961&0.6614\\
$(0,3,1)_r$-$(0,2,2)_r$ & 0.4764(42)&0.5373&0.6614\\
$(0,0,5)_r$-$(0,1,4)_r$ & 0.2957(35)&0.3134&0.6038\\
$(0,2,3)_r$-$(0,1,4)_r$ & 0.5187(75)&0.5161&0.7638\\

\hline
\end{tabular}
\end{center}
\label{tab:Cor_matrix}
\end{table}

\begin{table}[t]
\caption{C$_s$ equilibrium geometry obtained in the 3D fit. The number in parentheses after the $\theta$ value reflects the difference between the derived value of $\theta$ at the minima of the PES of the two isotopologues used in this study. The geometry derived from our 2D fit is reproduced here for comparison.}
\begin{center}
\begin{tabular}{@{\hspace{6pt}} c @{\hspace{6pt}} @{\hspace{6pt}} c @{\hspace{6pt}}  @{\hspace{6pt}} c   @{\hspace{6pt}} @{\hspace{6pt}}c  @{\hspace{6pt}}@{\hspace{6pt}} c  @{\hspace{6pt}}  @{\hspace{6pt}} c   @{\hspace{6pt}}c   @{\hspace{6pt}} c @{\hspace{6pt}}  @{\hspace{6pt}} c   @{\hspace{6pt}}c   @{\hspace{6pt}} c @{\hspace{6pt}} @{\hspace{6pt}} c @{\hspace{6pt}}}

\hline
&3D fit&\textit{ab initio}~\cite{Nachtigall1999}&2D fit\\ \hline
$r$$_1$/$\text{\AA}$ & 1.639&1.633&1.642\\
$r$$_2$/$\text{\AA}$ & 1.494&1.488&1.494\\
$\theta$/deg & 103.95(1)&103.3&103.80\\

\hline
\end{tabular}
\end{center}
\label{tab:3DGeometry}
\end{table}

As in the 2D fit, the C$_s$ equilibrium geometry is determined (Table~\ref{tab:3DGeometry}). The C$_s$ geometry agrees well with $ab$ $initio$ values~\cite{Nachtigall1999}. Recall that we constrained only the barrier height, $b$, and the shape parameter, $\rho$, to be isotopologue-independent. Even though we did not constrain the isotopologue independence of the absolute minimum geometry, the calculated minimum geometries given in Table~\ref{tab:3DGeometry} for S$^{16}$O$_2$ and S$^{18}$O$_2$ agree well with each other, which attests to the isotopologue independence of the PES. Note that the barrier height derived from our 3D fit differs by more than 10$\%$ from the 2D value (compare the values of $b$ in Table~\ref{tab:2D} and~\ref{tab:3D}). Without the ability to vary the shape parameter, $\rho$, the value of which is highly correlated with the barrier height, it is difficult to evaluate the accuracy of our derived barrier height.

\begin{figure}[b]
\includegraphics[width=3.5 in]{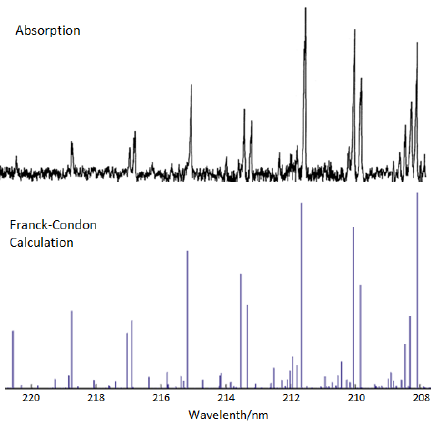}
\caption{Comparison of anharmonic Franck-Condon factors, calculated from the internal coordinate force field, with the low-resolution absorption spectrum (jet-cooled condition) in the 208-221 nm region. The experimental spectrum is adapted and reproduced with permission from Chem. Phys. Lett. \textbf{294}, 571 (1998). Copyright 1998 Elsevier. Even though none of the levels in the region with wavelength shorter than 220 nm are included directly in our fit, the calculated Franck-Condon intensities in this energy region agree well with the observed intensity pattern.}
\label{fig:fc_short}
\end{figure}

Experimental observables not included in the fit are well reproduced using our force field (see Tables~\ref{tab:3DEnergy16}, \ref{tab:3Drot}, and \ref{tab:Cor_matrix}). For the nine vibrational levels in the 3000--4000 cm$^{-1}$ region, which we do not include in our fit, the rms error between the calculated and the observed terms is 5 cm$^{-1}$ (rms=3.5 cm$^{-1}$, if we exclude the 3887.7 cm$^{-1}$ level). The Franck-Condon factors calculated from our force field also agree well with the observed intensity pattern in the absorption~\cite{{Freeman1984},{Hitran2012},{Sako1998}} and LIF~\cite{Yamanouchi1995} spectra. As shown in Fig.~\ref{fig:fc_short}, despite the fact that we have not input to the fit any of the vibrational term energies of levels with transition wavelength shorter than 220 nm (corresponding to levels $>$3000 cm$^{-1}$ above the $\tilde{\text{C}}$ state zero-point level), the calculated Franck-Condon intensities in this energy region agree well with the observed intensity patterns. The rotational constants that are not included in the fit (Table~\ref{tab:3Drot}) and the Coriolis matrix elements between highly anharmonic  $\tilde{\text{C}}$ state vibrational levels (Table~\ref{tab:Cor_matrix}) are also well reproduced from our force field (see detailed discussions in Section~\ref{subsec:Coriolis}).

\subsection{Vibrational Assignment Scheme}
\label{subsec:vib_assign}

\begin{figure}[b]
\includegraphics[width=5.5 in]{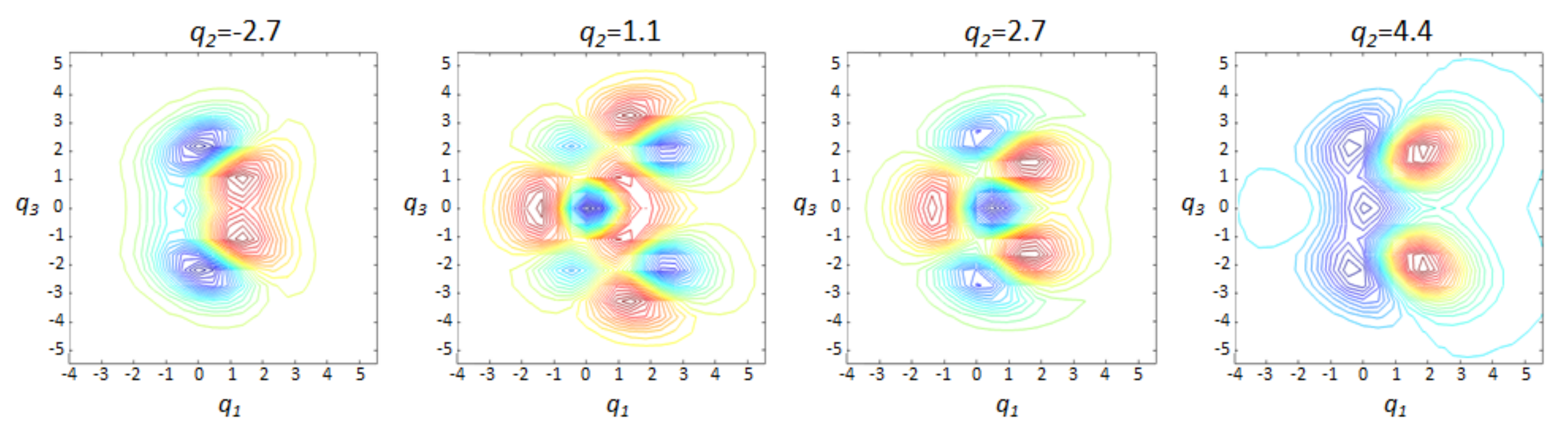}
\caption{Projections of the wavefunction of the 2394 cm$^{-1}$ state onto the $q_1$-$q_3$ plane at different values of $q_2$. The wavefunction is obtained directly from DVR calculations, using the force constants derived from the 3D fit.}
\label{fig:state_2394}
\end{figure}

How are the eigenstates from the 3D fit assigned? In the 2D case, Kellman-type vibrational assignments can be made easily based on visual inspection of the wavefunction, but visual assignment of the three-dimensional eigenstates is more challenging. Figure~\ref{fig:state_2394} illustrates the projections of the wavefunction of the 2394 cm$^{-1}$ state onto the $q_1$-$q_3$ plane at different values of $q_2$.  If the $\nu_2$ bending mode is rigorously separable from the stretching modes, the projection of the wavefunction should be independent of the value of $q_2$ (although the relative amplitude of each projection will depend on $q_2$, i.e.\ it will be small when the value of $q_2$ is near a node of the wavefunction along $q_2$ direction). This is, however, obviously not the case, as can be seen in Fig.~\ref{fig:state_2394}. Depending on which projection of the wavefunction we choose to look at, different Kellman-type vibrational assignments can be made. If we only consider basis states that are close in energy to the observed eigenstate, at $q_2=1.1$, the wavefunction could be assigned as (1,0,4)$_r$, which is predicted by our 2D model to occur at 2372 cm$^{-1}$. However, at $q_2$=2.7, the nodal pattern of the projection suggests $v_\alpha$=1 and $v_\beta$=2, consistent with an assignment (1,2,2)$_r$. The eigenstate in Fig. \ref{fig:state_2394} most likely contains contributions from both Kellman-type basis states. Therefore, we conclude that the separability of $\nu_2$ is not rigorous above 2000 cm$^{-1}$, and assignment schemes based on visual inspection of the 3D wavefunctions will be impractical.

To quantify the contributions from different Kellman-type basis states to each vibrational eigenstate, we transform our complete normal-mode Hamiltonian defined in Eq.~(\ref{eq:Hamiltonian}) into a Hamiltonian in a new Kellman basis. The new basis takes into account the two most prominent vibrational features of the $\tilde{\text{C}}$ state, i.e. strong interactions between $\nu_1$ and $\nu_3$ and a double-well in the $q_3$ coordinate. The bending mode $\nu_2$ is completely separable from the two stretching modes in the Kellman basis. In this work, the Kellman basis states are constructed via $\textit{partial}$ $\textit{diagonalization}$ of the original normal-mode Hamiltonian. Specifically, all the terms in Table~\ref{tab:3D} that involve interaction between $\nu_2$ and the other two modes are set to zero (e.g. $\phi_{233}$, $\phi_{122}$, $\phi_{2233}$ etc.), which results in a new Hamiltonian, $H_t$. A basis transformation, $V_t$, between the harmonic basis states and our Kellman basis states is obtained by diagonalizing $H_t$. The transformed Hamiltonian in the Kellman basis, $H_K$, is given by
\begin{equation}
H_K=V_t^{-1}HV_t,
\label{eq:Ht}
\end{equation}
where $H$ is the original Hamiltonian expressed in the harmonic basis. The $\phi_{222}$ and $\phi_{2222}$ terms are not set to zero in $H_t$, thus the intra-mode anharmonicity in $\nu_2$ is partially accounted for in our new basis set. Since $\nu_2$ is rigorously uncoupled from the other modes in this basis, we can easily assign ${definite}$ quantum numbers, $(v_\alpha,v_2,v_\beta)_{l/r}$, similarly defined as in Section~\ref{subsec:Fermi}, to the new basis states by visual inspection of each basis state wavefunction. Projections of the Kellman basis state wavefunctions onto the $q_1$\textendash$q_3$ plane are qualitatively similar to those shown in Fig.~\ref{fig:2Dpolyad}, and the quantum number, $v_2$, is obtained by counting the number of nodes along the $q_2$ coordinate. By diagonalizing the transformed Hamiltonian, we obtain the composition of each eigenstate as a linear combination of Kellman basis states. 

Using this two-step diagonalization method, the 2394 cm$^{-1}$ state has the following basis composition:
\begin{multline}
\ket{2394\,\,\text{cm}^{-1}}=\{0.8\mathbf{\ket{1,0,4}_r}+0.2\ket{1,1,4}_r\}-\{-0.05 \ket{1,1,2}_r+0.4 \mathbf{\ket{1,2,2}_r}+0.2 \ket{1,3,2}_r\} \\
+0.2\ket{0,1,6}_r+0.1\ket{2,1,0}_l+\cdot \cdot \cdot.
\label{eq:2394state}
\end{multline}
Equation~(\ref{eq:2394state}) shows that the 2394 cm$^{-1}$ state consists primarily of the Kellman $\ket{1,0,4}_r$ state, with an additional contribution from $\ket{1,2,2}_r$, in agreement with the result of visual deperturbation described earlier. Throughout this work, we use quantum numbers enclosed in parentheses, i.e. $(v_\alpha,v_2,v_\beta)_{r/l}$, to denote the vibrational assignment of an eigenstate (based on the dominant vibrational character of the state), and we reserve kets, i.e. $\ket{v_\alpha,v_2,v_\beta}_{r/l}$, to denote basis states. Occasionally, as in Eq.~(\ref{eq:2394state}), kets with the vibrational term energy of the state are also used to describe eigenstates. The presence of the $\ket{1,1,4}_r$ and $\ket{1,3,2}_r$ basis states in the composition of the 2394 cm$^{-1}$ eigenstate indicates large $intra$-$mode$ anharmonicity in $\nu_2$. Recall that $intra$-$mode$ anharmonicity in $\nu_2$ has already been partially accounted for in our Kellman basis by the inclusion of $\phi_{222}$ and $\phi_{2222}$ terms in $H_t$. However, additional contributions to $\nu_2$ anharmonicity arise from other terms omitted from $H_t$, such as $\phi_{112}$ and $\phi_{233}$. The basis states enclosed within each pair of curly brackets in Eq.~(\ref{eq:2394state}) can be considered collectively as an $anharmonic$ Kellman state, (e.g. $0.8 \ket{1,0,4}_r+0.2\ket{1,1,4}_r$ as an $anharmonic$ $\ket{1,0,4}_r$ state, which we label as $\mathcal{A}\ket{1,0,4}_{r}$, and $-0.05\ket{1,1,2}_r+0.4\ket{1,2,2}_r+0.2\ket{1,3,2}_r$ as $\mathcal{A}\ket{1,2,2}_{r}$), just as a Morse-oscillator eigenstate can be expressed as a linear combination of normal-mode basis states. Therefore, the 2394 cm$^{-1}$ state is largely a linear combination of anharmonic $\mathcal{A}\ket{1,0,4}_r$ and $\mathcal{A}\ket{1,2,2}_r$ states, with some small contributions from two other states from the second line of Eq.~(\ref{eq:2394state}). The projections of $\mathcal{A}\ket{1,0,4}_{r}$ onto the $q_1$-$q_3$ plane clearly remain unchanged from those of the original Kellman $\ket{1,0,4}_r$ basis, since they only involve contributions from progressions in $\nu_2$. In addition, for $\mathcal{A}\ket{1,0,4}_r$, the original Kellman basis states within each pair of curly brackets have the correct relative phases and magnitudes such that the number of nodes along the $q_2$ direction is zero, despite contribution from $\ket{1,1,4}_r$. This can be verified by inspecting the wavefunction of the $anharmonic$ state. A similar argument applies to $\mathcal{A}\ket{1,2,2}_{r}$. Thus, in $\mathcal{A}\ket{v_\alpha,v_2,v_\beta}_r$, $v_2$ should be taken as an $anharmonic$ quantum number, while the meanings of $v_\alpha$ and $v_\beta$ remain unchanged from those of the original Kellman basis. $Intra$-$mode$ anharmonicity becomes larger for states with more quanta of excitation in $\nu_2$ (compare the partitioning of states in the first and second pair of curly brackets in Eq.~(\ref{eq:2394state})). This is not surprising given that the size of the $\phi_{112}$ and $\phi_{233}$ matrix elements, which connect different Kellman basis states that differ by one quantum of $\nu_2$, increase as the quantum number increases. 

We must also point out that for the 2394 cm$^{-1}$ state, the direct interaction matrix element between the $\ket{1,0,4}_r$ and $\ket{1,2,2}_r$ states in the transformed Hamiltonian is small (1 cm$^{-1}$) compared to the energy difference of the two basis states ($\sim$13 cm$^{-1}$). This means, in order to obtain the mixing coefficients in Eq. (\ref{eq:2394state}), there must be an additional strong indirect interaction path between the $\ket{1,0,4}_r$ and $\ket{1,2,2}_r$ states. In fact, $\ket{1,1,4}_r$ and $\ket{1,3,2}_r$ states act as the dominant intermediate states of the indirect coupling. For example, anharmonicity in $\nu_2$ connects $\ket{1,0,4}_r$ with $\ket{1,1,4}_r$. The matrix element between $\ket{1,1,4}_r$ and $\ket{1,2,2}_r$, 44 cm$^{-1}$, has appreciable magnitude. As a result, the $\ket{1,0,4}_r$ and $\ket{1,2,2}_r$ states interact indirectly via $\ket{1,1,4}_r$ and similarly via $\ket{1,3,2}_r$, or to put it in another way, the anharmonic $\mathcal{A}\ket{1,0,4}_r$ and $\mathcal{A}\ket{1,2,2}_r$ states, which have a larger effective matrix element, interact directly to give rise to the eigenstate at 2394 cm$^{-1}$. 

\begin{figure}[t]
\includegraphics[width=4.7 in]{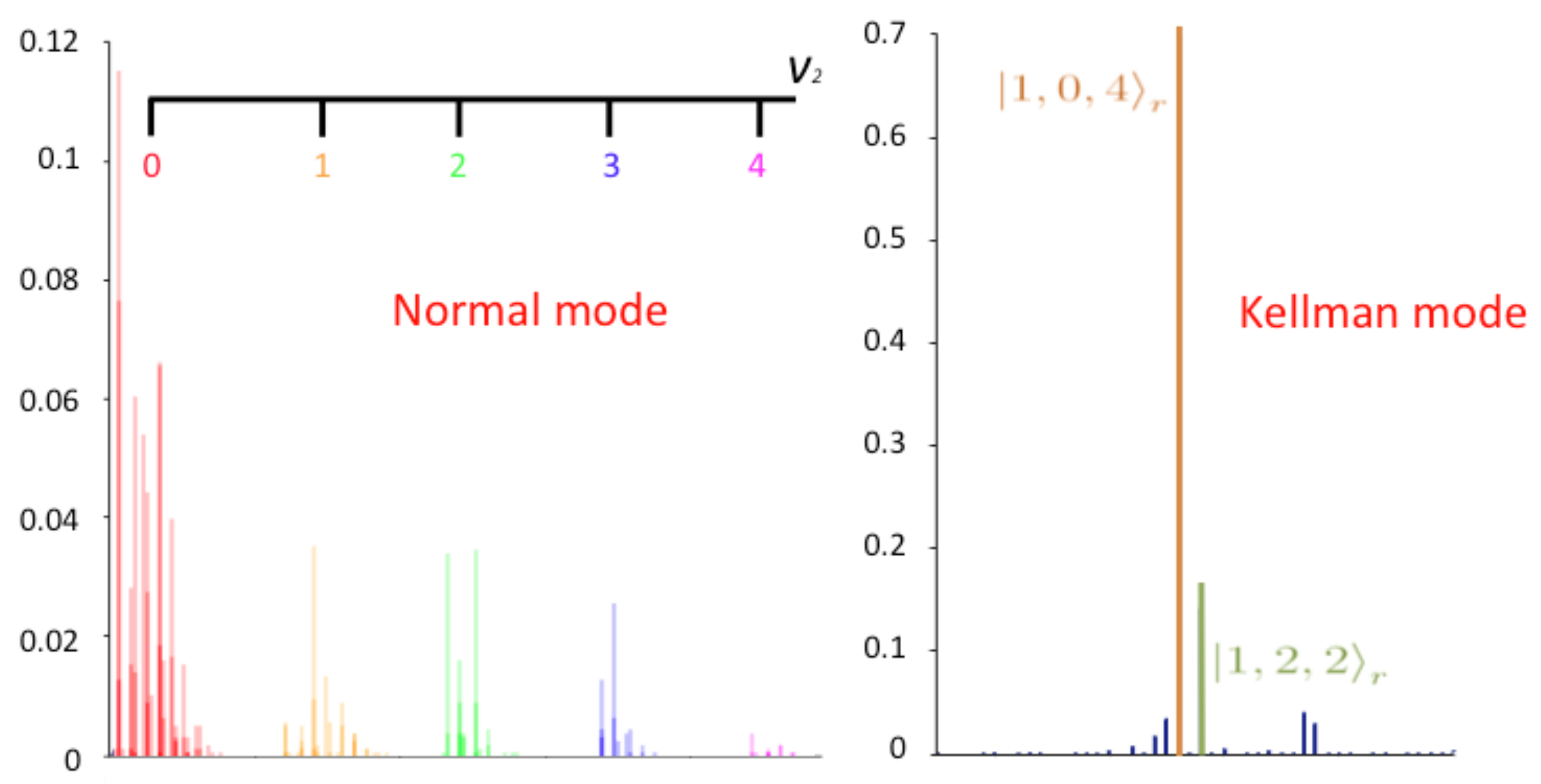}
\caption{Basis state distribution of the eigenstate at 2394 cm$^{-1}$ in normal-mode (left panel) and Kellman-mode (right panel) representations. The vertical axis represents the squares of the coefficient of a specific basis state. (See text for an explanation of the ordering of states in the normal-mode representation.)}
\label{fig:normal_kellman}
\end{figure}

Figure~\ref{fig:normal_kellman} displays the basis state distributions of the eigenstate at 2394 cm$^{-1}$ in both normal-mode and Kellman-mode representations. In the normal-mode representation shown on the left, basis states within each color-coded cluster belong to one value of $v_2$ (color-coded accordingly), and within each cluster, basis states are ordered according to quantum number $v_1$ followed by $v_3$. That means the first cluster contains harmonic basis states with $v_2=0$ arranged in the order of $\ket{0,0,0},\ket{0,0,2}...\ket{1,0,0},\ket{1,0,2}...\ket{2,0,0},\ket{2,0,2}...$ It is clear that the eigenstate character is more broadly distributed among basis states in the normal-mode representation than in the Kellman-mode representation. While it is impossible to identify a dominant basis state in the normal-mode representation, using the Kellman-mode representation, we are immediately able to identify the presence of two dominant basis states in the composition of the eigenstate. 

In Ref ~\onlinecite{Sako1998}, the vibrational eigenstates of the $\tilde{\text{C}}$ state were assigned based on visual inspection of the nodal patterns of the integrated 3D vibrational wavefunctions in the $q_1$-$q_3$ plane. The effects from the stretch-bend couplings were averaged over $q_2$ in their assignment scheme. The 2394 cm$^{-1}$ state was assigned as a pure Kellman-type (1,0,4)$_r$ state (translated into our notation). While this assignment is nominally correct, our analysis clearly shows that the 2394 cm$^{-1}$ state also has appreciable $\ket{1,2,2}_r$ character. Vibrational assignment based on visual inspection of the integrated wavefunction is flawed due to anharmonic interactions involving $\nu_2$. In contrast, we must emphasize that the vibrational character of each eigenstate can be unambiguously identified in our work from the eigenstate composition of the transformed Hamiltonian in the Kellman basis. 

\subsection{Franck-Condon Interference Effects}
\label{subsec:fc_int}

\begin{table}[b]
\caption{Estimated vibrational term energies and Franck-Condon factors (fc), relative to the (0,0) transition, for three states near 2750 cm$^{-1}$, calculated from the 2D internal force field.}
\begin{center}
\begin{tabular}{@{\hspace{6pt}} c @{\hspace{6pt}} @{\hspace{6pt}} c @{\hspace{6pt}}  @{\hspace{6pt}} c   @{\hspace{6pt}}c  @{\hspace{6pt}} c  @{\hspace{6pt}}  @{\hspace{6pt}} c   @{\hspace{6pt}}c   @{\hspace{6pt}} c @{\hspace{6pt}}  @{\hspace{6pt}} c   @{\hspace{6pt}}c   @{\hspace{6pt}} c @{\hspace{6pt}} @{\hspace{6pt}} c @{\hspace{6pt}}}

\hline
Assig. & $E$/cm$^{-1}$ & fc   \\
\hline
(2,0,2)$_r$ & 2728 & 26\\
(1,1,4)$_r$  & 2750 & 41\\
(1,3,2)$_r$  & 2785 & 10\\

\hline
\end{tabular}
\end{center}
\label{tab:2Dfc}
\end{table}

Our 2D internal force field, discussed in Section~\ref{sec:2D fit}, provides the first evidence of Franck-Condon interference effects in the $\tilde{\text{C}}$ state of SO$_2$. Before presenting a more quantitative demonstration of the interference effects based on our 3D force field, we first give a brief discussion of the result from our 2D model. Using a constant $\omega_2$ frequency of 378 cm$^{-1}$, our 2D force field predicts three states, (1,1,4)$_r$, (2,0,2)$_r$ and (1,3,2)$_r$, near 2750 cm$^{-1}$, based on the calculated vibrational term energies of the corresponding states with $v_2=0$ in Table~\ref{tab:2Denergy}, e.g. (1,0,2)$_r$ is predicted at 1651 cm$^{-1}$, which puts (1,3,2)$_r$ at 2785 cm$^{-1}$. Given the calculated Franck-Condon intensities into the three corresponding $v_2=0$ levels from the 2D force field (see Fig.~\ref{fig:2Dpolyad}) and the observed $(0,v_2,0)$ Franck-Condon progression~\cite{{Yamanouchi1995},{Freeman1984},{Hitran2012}}, relative Franck-Condon factors can be estimated for (1,1,4)$_r$, (2,0,2)$_r$ and (1,3,2)$_r$. As can be seen from Table~\ref{tab:2Dfc}, all of the three close-lying levels obtained from the model have large Franck-Condon factors. However, there is only one strong transtion observed in this energy region at 2762 cm$^{-1}$~\cite{Yamanouchi1995}. Our Franck-Condon calculation, based on our 3D force field, also predicts only one strong transition in this energy region, along with two other much weaker transitions. Therefore, there must be interactions among the three Kellman-type levels. The interactions are capable of mixing levels that have different quanta of excitation in $\nu_2$, and they are sufficiently strong to cause nearly complete annihilation of the Franck-Condon intensities to two of the three states. 

Using our transformed Hamiltonian in the Kellman basis, the interference effect can be analyzed in more detail based on our 3D internal force field. The three eigenstates of interest have the following Kellman basis composition: 
 \begin{multline}
\ket{2729\,\, \text{cm}^{-1}}=\{-0.2{\ket{1,0,4}_r}+0.5\mathbf{\ket{1,1,4}_r}+0.3{\ket{1,2,4}_r}\}+\{0.4\mathbf{\ket{2,0,2}_r}+0.1{\ket{2,1,2}_r}\}\\
\,\,\,\,\,\,\,\,\,\,\,\,\,\,\,\,\,\,\,\,\,\,\,\,\,\,\,\,\,\,\,\,\,\,\,\,\,\,\,\,\,\,\,\,\,\,\,\,\,\,\,\,\,\,\,\,\,\,\,\,\,\,-\{-0.3 \ket{1,2,2}_r+0.4\mathbf{\ket{1,3,2}_r}+0.2\ket{1,4,2}\}+\cdot\cdot\cdot \\ \,\,\,\,\ket{2744\,\,\text{cm}^{-1}}=\{{-0.06\ket{1,0,4}_r}+0.2\mathbf{\ket{1,1,4}_r}+0.01{\ket{1,2,4}_r}\}+\{0.6\mathbf{\ket{2,0,2}_r}+0.2\ket{2,1,2}_r\}\\
\,\,\,\,\,\,\,\,\,\,\,\,\,\,\,\,\,\,\,\,\,\,\,\,\,\,\,\,\,\,\,\,\,\,\,\,\,\,\,\,\,\,\,\,\,\,\,\,\,\,\,\,\,\,\,\,\,\,\,\,\,\,+\{-0.2 \ket{1,2,2}_r+0.5\mathbf{\ket{1,3,2}_r}+0.3\ket{1,4,2}\}+\cdot\cdot\cdot \\
\,\,\,\,\ket{2762\,\, \text{cm}^{-1}}=\{-0.2 \ket{1,0,4}_r+0.6\mathbf{\ket{1,1,4}_r}+0.2\ket{1,2,4}_r\}-\{0.6 \mathbf{\ket{2,0,2}_r}+0.1{\ket{2,1,2}_r}\}\\
+\{-0.02{\ket{1,2,2}_r}+0.3 \mathbf{\ket{1,3,2}_r}+0.2{\ket{1,4,2}_r}\}+\cdot\cdot\cdot
\label{eq:2750states}
\end{multline}
As in Eq.~(\ref{eq:2394state}), basis states enclosed in each pair of curly brackets can be considered collectively as an $anharmonic$ state. The anharmonic $\mathcal{A}\ket{1,1,4}_r,\mathcal{A}\ket{2,0,2}_r,\mathcal{A}\ket{1,3,2}_r$ states account for about $90\%$ of the total composition of the three eigenstates in Eq.~(\ref{eq:2750states}). Note that basis states in each pair of curly brackets have consistent relative phases and approximately consistent amplitudes such that the $anharmonic$ $\nu_2$ quantum number is meaningful. Given that the 2762 cm$^{-1}$ state predominantly consists of Kellman basis states with $v_2=0$ or 1, the widely-adopted assignment of the 2762 cm$^{-1}$ state as a normal-mode (1,3,2) level~\cite{{Yamanouchi1995},{Bludsky2000},{Xie1999}}, based on an apparent $(1,n,2)$ progression, is $incorrect$, even when we interpret the assignment as Kellman (1,3,2)$_r$. 

\begin{table}[b]
\caption{Franck-Condon factors, relative to that of the (0,0) transition, calculated from the three-state interaction model described in the text. The contributions from each $anharmonic$ Kellman state to the vibrational overlap integral of the three eigenstate wavefunctions with the $\tilde{\text{X}}$ state zero-point level are calculated using the three-state model, and the results are listed in columns 2--4. Vibrational overlap integrals and Franck-Condon factors of the three `eigenstates' in the three-state interaction model are then calculated and listed respectively in the overlap and fc$_{model}$ column. Franck-Condon factors for the actual eigenstates calculated from a full 3D calculation are listed in the fc column for comparison. Note that, experimentally, transitions into the 2762 cm$^{-1}$ level are about 200--300 times stronger than transitions into the 2743 cm$^{-1}$ level (see Fig. 5 in Ref ~\onlinecite{Park2015}). The 2730 cm$^{-1}$ level has not been experimentally observed.}
\begin{center}
\begin{tabular}{@{\hspace{6pt}} c @{\hspace{6pt}} @{\hspace{6pt}} c @{\hspace{6pt}}  @{\hspace{6pt}} c   @{\hspace{6pt}} @{\hspace{6pt}}c  @{\hspace{6pt}}@{\hspace{6pt}} c  @{\hspace{6pt}}  @{\hspace{6pt}} c   @{\hspace{6pt}}c   @{\hspace{6pt}} c @{\hspace{6pt}}  @{\hspace{6pt}} c   @{\hspace{6pt}}c   @{\hspace{6pt}} c @{\hspace{6pt}} @{\hspace{6pt}} c @{\hspace{6pt}}}

\hline
&$\mathcal{A}\ket{1,1,4}_r$&$\mathcal{A}\ket{2,0,2}_r$&$\mathcal{A}\ket{1,3,2}_r$&overlap&fc$_{model}$&fc\\ \hline
$\ket{2730\,\, \text{cm}^{-1}}$ & 3.1&	-2.1&-0.8&0.2&0.04&0.5\\
$\ket{2743\,\, \text{cm}^{-1}}$ & -0.4&	3.4&-1.6&1.4&1.9&0.4\\
$\ket{2762\,\, \text{cm}^{-1}}$ & 3.3&	2.6&1.2&7.1&50&82\\

\hline
\end{tabular}
\end{center}
\label{tab:interference}
\end{table}

To demonstrate the interference effect on the Franck-Condon intensities of the three eigenstates in Eq. (\ref{eq:2750states}), we consider only the three anharmonic states, $\mathcal{A}\ket{1,1,4}_r,\mathcal{A}\ket{2,0,2}_r$, and $\mathcal{A}\ket{1,3,2}_r$. The contribution from each anharmonic Kellman state to the vibrational overlap integral of an eigenstate in Eq. (\ref{eq:2750states}) with the $\tilde{\text{X}}$ state zero-point wavefunction is calculated, using the basis transformation matrix, $V_t$, in Eq. (\ref{eq:Ht}). Franck-Condon factors for each `eigenstate' in the three-state interaction model can then be calculated and these are summarized in Table~\ref{tab:interference}. Our three-state model indeed reproduces the experimental observation that there is only one strong transition at 2762 cm$^{-1}$. Vibrational overlap integrals of the three anharmonic states combine constructively for the 2762 cm$^{-1}$ eigenstate, but almost perfectly cancel for the other two eigenstates. This causes near-annihilation of Franck-Condon intensity in two of the three states and an enhancement for the third state. Such Franck-Condon interference effects are prevalent for vibrational levels above 2500 cm$^{-1}$. This indicates a serious breakdown of the assumption of separability of the bending motion from the other two strongly interacting motions, especially for states above 2500 cm$^{-1}$. The Kellman basis states are no longer sufficient to describe the dynamics in this energy region. Interference effects like these suggest the emergence of a new class of zero-order state. Decoding the new dynamics poses an interesting challenge for future work. In Tables~\ref{tab:3DEnergy16} and \ref{tab:3Denergy_rest}, we label the eigenstates according to the degree of interaction in the Kellman basis. For some of the levels, no vibrational assignment can be given, since none of the $anharmonic$ Kellman basis states has greater than $50\%$ of the character of the eigenstate. 

\subsection{Rotational Information and Vibrational Assignments}
\label{subsec:rotational}

Many of our vibrational assignments are confirmed by the magnitude of the experimentally-derived rotational constants, especially the $A$ rotational constants. The $A$ constant increases by $\sim$0.02 cm$^{-1}$ per quantum of excitation of the bending mode, compared to $\sim$0.003 cm$^{-1}$ for $\nu_1$ and $\nu_3$. Thus, the number of quanta in $\nu_2$ can be estimated qualitatively from $v_2\approx (A_v-A_0)/(0.02$ cm$^{-1}$), where $A_0$=1.1505 cm$^{-1}$ is the $A$ constant of the $\tilde{\text{C}}$ state (0,0,0)$_r$ level. Here, we demonstrate the use of rotational information to provide an additional check on the vibrational assignment of the 2762 cm$^{-1}$ level, which is discussed in Section~\ref{subsec:fc_int}.

The level at 2762 cm$^{-1}$ is rotationally perturbed by strong $c$-axis Coriolis interactions with nearby $b_2$ vibrational levels, leading to large uncertainties in the effective rotational constants. For example, the effective A constant of the 2762 cm$^{-1}$ level is $1.186\pm0.044$ cm$^{-1}$ (2$\sigma$ uncertainty). This uncertainty is sufficiently large that it precludes determination of the number of quanta of excitation in $\nu_2$. Recently, we implemented the Coherence-Converted Population Transfer technique\cite{Pate2010} in a sensitive, background-free scheme for millimeter-wave optical double resonance (CCPT-MODR)~\cite{Park2015} to probe the vibrational levels near the 2762 cm$^{-1}$ state, including one dark $b_2$  symmetry level, which borrows intensity via the Coriolis interaction. We assigned 16 rotational levels of the dark $b_2$ symmetry level (as well as many additional rotational levels of the 2762 cm$^{-1}$ state and 4 rotational levels of an $a_1$ symmetry level at 2743 cm$^{-1}$). This allowed us to deperturb the Coriolis interactions among the three observed vibrational states. The deperturbed $A$ rotational constant of the 2762 cm$^{-1}$ level is 1.169(7) cm$^{-1}$, which indicates that this level effectively has approximately one quantum of excitation in $\nu_2$, in agreement with our analysis in Section~\ref{subsec:fc_int} indicating that the 2762 cm$^{-1}$ level is predominately a linear combination of $\mathcal{A}\ket{1,1,4}_r$ and $\mathcal{A}\ket{2,0,2}_r$, with a smaller contribution from $\mathcal{A}\ket{1,3,2}_r$. The magnitude of the newly derived $A$ constant helps us to rule out the original vibrational assignment of this state as (1,3,2)~\cite{{Yamanouchi1995},{Bludsky2000},{Xie1999}}, which should have an $A$ constant close to 1.20 cm$^{-1}$. 

We can compare the rotational constants of the three states studied in the CCPT-MODR experiments to our calculated rotational constants (see Table~\ref{tab:MODR}). Note that the three states are labeled according to the notations used in our CCPT-MODR paper on the $\tilde{\text{C}}$ state~\cite{Park2015}, where the label `B' is given for the Franck-Condon bright state at 2762 cm$^{-1}$ and `P' for the two Franck-Condon dark perturbing eigenstates. The subscripts indicate the vibrational symmetry of the levels. The calculated $A$ constant for the 2762 cm$^{-1}$ (B$_{\text{a}_1}$) level in Table VII falls very close to the $2\sigma$ uncertainty of the experimentally derived value, which supports the accuracy of our eigenstate expansion. The experimental and calculated $A$ constants for the 2753 cm$^{-1}$ (P$_{\text{b}_2}$) level suggest $v_2$=0, consistent with the vibrational assignment (1,0,5)$_r$.  

Note that only the first eight $b_2$ vibrational symmetry levels have been experimentally observed in the IR-UV double resonance experiment~\cite{Park2015SO2}. This means that there is a large energy gap between the last observed $b_2$ level at 1595 cm$^{-1}$ and this level at 2753 cm$^{-1}$. However, we believe that we match the eigenstate correctly because, according to our calculation, there are only four $b_2$ symmetry levels between 2700 cm$^{-1}$ and 2800 cm$^{-1}$, and only one of these has no excitation in $\nu_2$. The other three levels have at least two quanta in $\nu_2$. Using this eigenstate assignment, the calculated Coriolis matrix element, $t_1$, between the B$_{\text{a}_1}$ and P$_{\text{b}_2}$ levels in Table~\ref{tab:MODR} agrees with the experimental value. The assignment of the other $a_1$ level at 2743 cm$^{-1}$ is less certain, given that there are two calculated $a_1$ levels in the small energy region, 2740\textendash2745 cm$^{-1}$. In addition, only four rotational term energies belonging to this level have been observed~\cite{Park2015}, resulting in large uncertainty in its molecular constants. However, the P$_{\text{a}_1}$ level most likely corresponds to the calculated level at 2744.5 cm$^{-1}$ in Table~\ref{tab:3DEnergy16} because, using this assignment, the calculated value of the Coriolis matrix element is in better agreement with the experimental value. In addition, the experimental $A$ rotational constant of the P$_{\text{a}_1}$ level, although not precisely determined, indicates moderate excitation in $\nu_2$, allowing us to rule out the (0,6,2)$_r$ level predicted at 2741 cm$^{-1}$ (Table~\ref{tab:3DEnergy16}), which has substantial excitation in $\nu_2$. However, due to ambiguity in eigenstate assignment for the observed state at 2743 cm$^{-1}$, the observed P$_{\text{a}_1}$ level in Table~\ref{tab:MODR} is not included in our fit.

\begin{table}[b]
\caption{Calculated ($A_{\text{c}}$) and experimentally derived ($A_{\text{o}}$) rotational $A$ constants and Coriolis matrix elements of three states near 2762 cm$^{-1}$ observed in MODR study. For convenience, each state is given the label assigned in Ref ~\onlinecite{Park2015}.}
\begin{center}
\begin{tabular}{@{\hspace{0pt}} c @{\hspace{0pt}} @{\hspace{3pt}} c @{\hspace{6pt}}  @{\hspace{6pt}} c   @{\hspace{6pt}} @{\hspace{6pt}}c  @{\hspace{6pt}}@{\hspace{6pt}} c  @{\hspace{6pt}}  @{\hspace{6pt}} c   @{\hspace{6pt}}c   @{\hspace{6pt}} c @{\hspace{6pt}}  @{\hspace{6pt}} c   @{\hspace{6pt}}c   @{\hspace{6pt}} c @{\hspace{6pt}} @{\hspace{6pt}} c @{\hspace{6pt}}}

\hline
&E/cm$^{-1}$&Assig.&$A_{\text{o}}$/cm$^{-1}$~\cite{Park2015}&$A_\text{c}$/cm$^{-1}$\\
\hline
B$_{\text{a}_1}$&2762 & &1.169(7)& 1.159 \\
P$_{\text{b}_2}$&2753& (1,0,5)$_r$ &1.1410(10)&	1.1388\\
P$_{\text{a}_1}$&2743&   &1.09(10)&	1.18\\ \hline
\multicolumn{2}{c}{}&\multicolumn{1}{c}{Exp./cm$^{-1}$}&\multicolumn{1}{c}{Cal./cm$^{-1}$}\\ 
\hline
\multicolumn{2}{c}{$t_1(\text{B}_{\text{a}_1},\text{P}_{\text{b}_2})$}&\multicolumn{1}{c}{0.43(4)}&\multicolumn{1}{c}{0.45} \\
\multicolumn{2}{c}{$t_1(\text{P}_{\text{b}_2},\text{P}_{\text{a}_1})$}&\multicolumn{1}{c}{0.15(4)}&\multicolumn{1}{c}{0.19} \\

\hline
\end{tabular}
\end{center}
\label{tab:MODR}
\end{table}

Above 3000 cm$^{-1}$, very few vibrational levels of the $\tilde{\text{C}}$ state have been observed in the high-resolution LIF study from Yamanouchi et al.~\cite{Yamanouchi1995}, because of predissociation. Moreover, vibrational level density in the 3000\textendash4000 cm$^{-1}$ region is twice that between 2000\textendash3000 cm$^{-1}$. However, we believe that the vibrational assignments that are listed for the nine observed levels between 3000\textendash4000 cm$^{-1}$ (Table~\ref{tab:3DEnergy16}) are correct, because the assigned $v_2$ quantum numbers of those levels are consistent with the magnitudes of the experimental $A$ rotational constants. We emphasize that the assignment scheme developed in this work provides unambiguous vibrational assignments to an unprecedented number of the $\tilde{\text{C}}$ state levels, all consistent with available rotational information.

\subsection{The Coriolis Effects in the $\tilde{\text{C}}$ state of SO$_2$}
\label{subsec:Coriolis}

As recognized in earlier studies of the SO$_2$ $\tilde{\text{C}}$ state~\cite{{Brand1976},{Johann1977},{Brand1978},{Yamanouchi1995}}, the $C$ rotational constants are severely perturbed by $c$-axis Coriolis interactions between $\nu_2$ and $\nu_3$ (more precisely, $\nu_\beta$, due to Fermi-133 resonance). However, accurate deperturbation had been impossible until the recent direct observation of $b_2$ vibrational levels~\cite{{Park2015SO2},{Park2015}}. In this section, we present level-specific Coriolis interaction strengths derived from our force field, and we analyze how the double-well structure of the PES leads to specific diagnostic patterns of Coriolis interactions in the $\tilde{\text{C}}$ state.

First, we discuss the calculated rotational constants in Table~\ref{tab:3Drot}, in particular the $C$ rotational constants. The Coriolis contributions to the values of the Coriolis-$perturbed$ $C$ rotational constants are listed in the $C_{Cor}$ column of Table~\ref{tab:3Drot}. It is obvious that Coriolis contributions are significant (in some cases, $50\%$ contribution) to the values of the $C$ constants of the majority of vibrational levels. In addition, the sign of the Coriolis contribution indicates the relative locations of the interacting states~\cite{{Brand1978},{Johann1977},{Yamanouchi1995}}. A positive (negative) Coriolis contribution to the $C$ constant indicates that an interacting level lies below (above) the level of interest. The calculated $C_p$ constants agree well with the experimentally derived ones, except in the case of a few close-lying Coriolis-interacting pairs of levels. For example, the calculated $C$ constants for the (0,0,4)$_r$, and (0,1,3)$_r$ levels, whose vibrational origins are separated by only 7 cm$^{-1}$, differ from the experimentally derived values by 0.03 cm$^{-1}$. Note that the Coriolis contribution to the $C$ constants of these strongly perturbed levels is about 30\textendash40$\%$. It is not surprising that our nondegenerate perturbation treatment of the Coriolis interactions (Eq. (\ref{eq:Cor})) fails for those close-lying levels.

The calculated $A$, $B$, and $C_{dp}$ rotational constants, and the Coriolis matrix elements between vibrational levels calculated from Eq.~(\ref{eq:me}) (Table~\ref{tab:Cor_matrix}), can be compared directly with deperturbed rotational constants and Coriolis matrix elements reported in the first paper of this series~\cite{Park2015SO2}. For some levels, deperturbed $C$ constants are not available from the experiment, because the Coriolis-interacting states are distant in energy and Coriolis deperturbation is not possible without observations of high-$J$ levels. The agreement between the experiment and our calculation is good for most cases, considering that it is unclear whether the experimentally derived values are fully deperturbed. (Due to lack of high-$J$ data points for most of the observed vibrational levels, approximations must be made in the Coriolis fit~\cite{Park2015SO2}, in order to reduce correlations among the fit parameters. However, the general agreement between the experimental and calculated values supports the validity of these approximations.) As can be seen in Table~\ref{tab:Cor_matrix}, the calculated and experimental Coriolis matrix elements are both smaller than the harmonic predictions by $\sim$20--50$\%$. In the first part of this series~\cite{Park2015SO2}, this decrease in the effective Coriolis interactions is explained in terms of anharmonic effects (Fermi-133 and Darling-Dennison-1133). Our calculated Coriolis matrix elements, which explicitly take into account anharmonic effects in the $\tilde{\text{C}}$ state vibrational levels, successfully reproduce this decrease in the effective Coriolis interactions due to anharmonic interactions.

Just as Fermi-133 resonance is unusual in symmetric triatomic molecules, strong Coriolis mixing between the bending ($\nu_2$) and antisymmetric stretching ($\nu_3$) modes is also unusual, because the $\nu_2$ and $\nu_3$ modes are not typically close in frequency. These resonances occur in the $\tilde{\text{C}}$ state of SO$_2$ because the double-well dramatically depresses the effective $\nu_3$ frequency. To demonstrate the structure of Coriolis interactions in the $\tilde{\text{C}}$ state of SO$_2$, the rotationless Coriolis mixing angles between the two states in question, defined as $|\frac{t_1}{\Delta E}|$, where $\Delta E$ is the energy difference between the band-origins of the two states, are displayed in Fig.~\ref{fig:CorPolyad}. The mixing angles, which measure the extent of Coriolis interactions, are taken to be positive and are color-coded in Fig.~\ref{fig:CorPolyad}. In the absence of indirect higher-order interaction, $a_1$ vibrational symmetry levels interact only with $b_2$ symmetry levels via Coriolis interaction. 

In Fig.~\ref{fig:CorPolyad}, Coriolis-interacting states are grouped together to indicate that they form a Coriolis polyad (designated by a polyad label, $P_n$, where $n=v_2+v_\beta$). Vibrational levels within each Coriolis polyad interact strongly, while inter-polyad interactions are much weaker. Note that even if our assignments are based on Kellman's semiclassical assignment scheme, the selection rule for Coriolis interactions appears to be very similar in form to the harmonic selection rule. In the harmonic case, when $\nu_2$ and $\nu_3$ have similar frequencies, $(0,v_2,v_3-1)$ and $(0,v_2+1,v_3)$ are a pair of Coriolis interacting states, while in our assignment scheme, $(0,v_2,v_\beta-1)_r$ and $(0,v_2+1,v_\beta)_r$ are strongly Coriolis interacting (when $v_\beta>1$). For levels that lie below 1600 cm$^{-1}$ in the $\tilde{\text{C}}$ state, where $all$ of the $a_1$ and $b_2$ symmetry levels have been experimentally observed, this selection rule seems to be obeyed and the Coriolis polyads are formed based on this selection rule, as is schematically displayed in Fig.~\ref{fig:CorPolyad}. 

\begin{figure}[t]
\includegraphics[width=3.5 in]{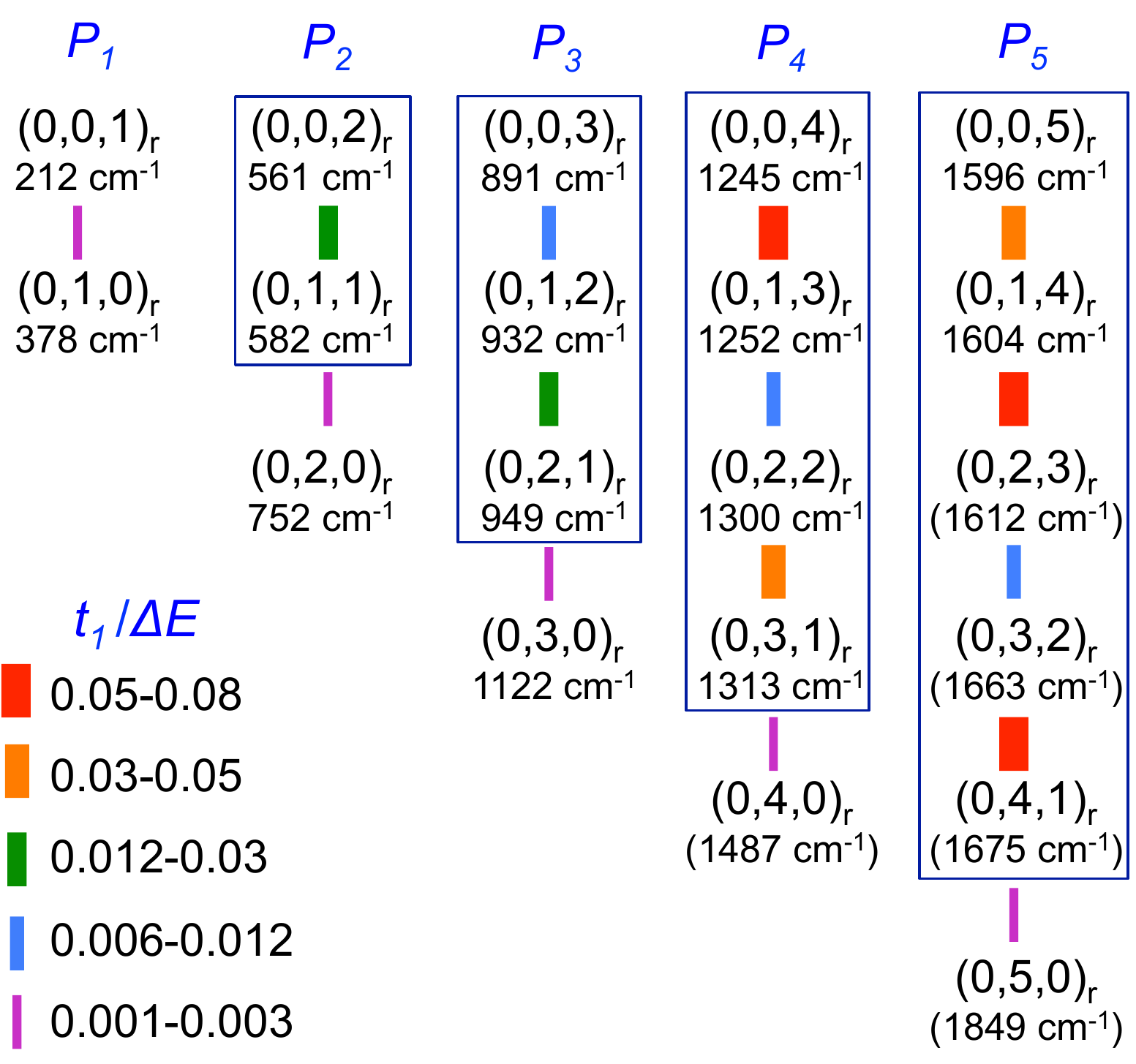}
\caption{Formation of Coriolis polyads. Note that energies of the vibrational states increase from left to right and from top to bottom. The lines between different pairs of states are drawn with different color and width, in order to depict the magnitude of the rotationless Coriolis mixing angle. Strongly interacting states forming a polyad (designated by the polyad number, $P_n$) are grouped together. The calculated vibrational term energies are used for levels not observed experimentally and are enclosed in parentheses.} 
\label{fig:CorPolyad}
\end{figure}

Note that the mixing-angles of pairs of levels in the $P_1$--$P_5$ polyads in Fig.~\ref{fig:CorPolyad} seem to develop an alternating pattern, i.e.\ the mixing angle of the $(0,v_2,v_\beta)$-$(0,v_2+1,v_\beta-1)$ pair does not increase monotonically as $v_\beta$ increases, and the mixing angle between the $a_1$ and $b_2$ symmetry levels within a $P_n$ polyad oscillates as one moves down the $P_n$ column in Fig.~\ref{fig:CorPolyad}.  As is evident from Table~\ref{tab:Cor_matrix}, both the experiment and the calculation based on our force field show that the Coriolis matrix elements, $t_1$, are similar in this energy region. \textit{Thus, the variation in the mixing angles in Fig.~\ref{fig:CorPolyad} is mostly due to the variation of the energy difference between the two states in question.} The alternating pattern in the Coriolis mixing angle is in fact a manifestation of the effects of the double-well on the PES on the rotational structure of the molecule. 

To see how the double-well structure leads to the alternating pattern, we define the effective $\nu_\beta$ frequency of the $(0,v_2,v_\beta)_r$ level as the energy difference between the $(0,v_2,v_\beta)_r$ and $(0,v_2,v_\beta+1)_r$ levels. Similarly, the effective $\nu_2$ frequency of $(0,v_2,v_\beta)_r$ is the energy difference between $(0,v_2,v_\beta)_r$ and $(0,v_2+1,v_\beta)_r$.  The effective $\nu_2$ frequency is approximately a constant ($\sim$ 377 cm$^{-1}$). Due to the double-well structure of the PES, the magnitude of the effective $\nu_\beta$ frequency alternates as a function of $v_\beta$, as can be seen in Fig.~\ref{fig:zigzag}. The effective $\nu_\beta$ frequency of $(0,v_2,v_\beta)_r$ is larger than that of $(0,v_2',v_\beta+1)_r$ if $v_\beta$ is odd, and it is smaller than that of $(0,v_2',v_\beta+1)_r$ if $v_\beta$ is even. Given this alternating pattern in the effective $\nu_\beta$ frequency, the alternations in the mixing angles down each of the $P_2$--$P_5$ polyad columns in Fig.~\ref{fig:CorPolyad} can be explained. Similar arguments are applicable to the oscillating patterns across each row in Fig.~\ref{fig:CorPolyad}. As shown here, the staggering in the vibrational energy spacings results in an alternation in the degree of Coriolis interactions between levels in the $P_1$--$P_5$ polyads. 

\begin{figure}[b]
\includegraphics[width=3.1 in]{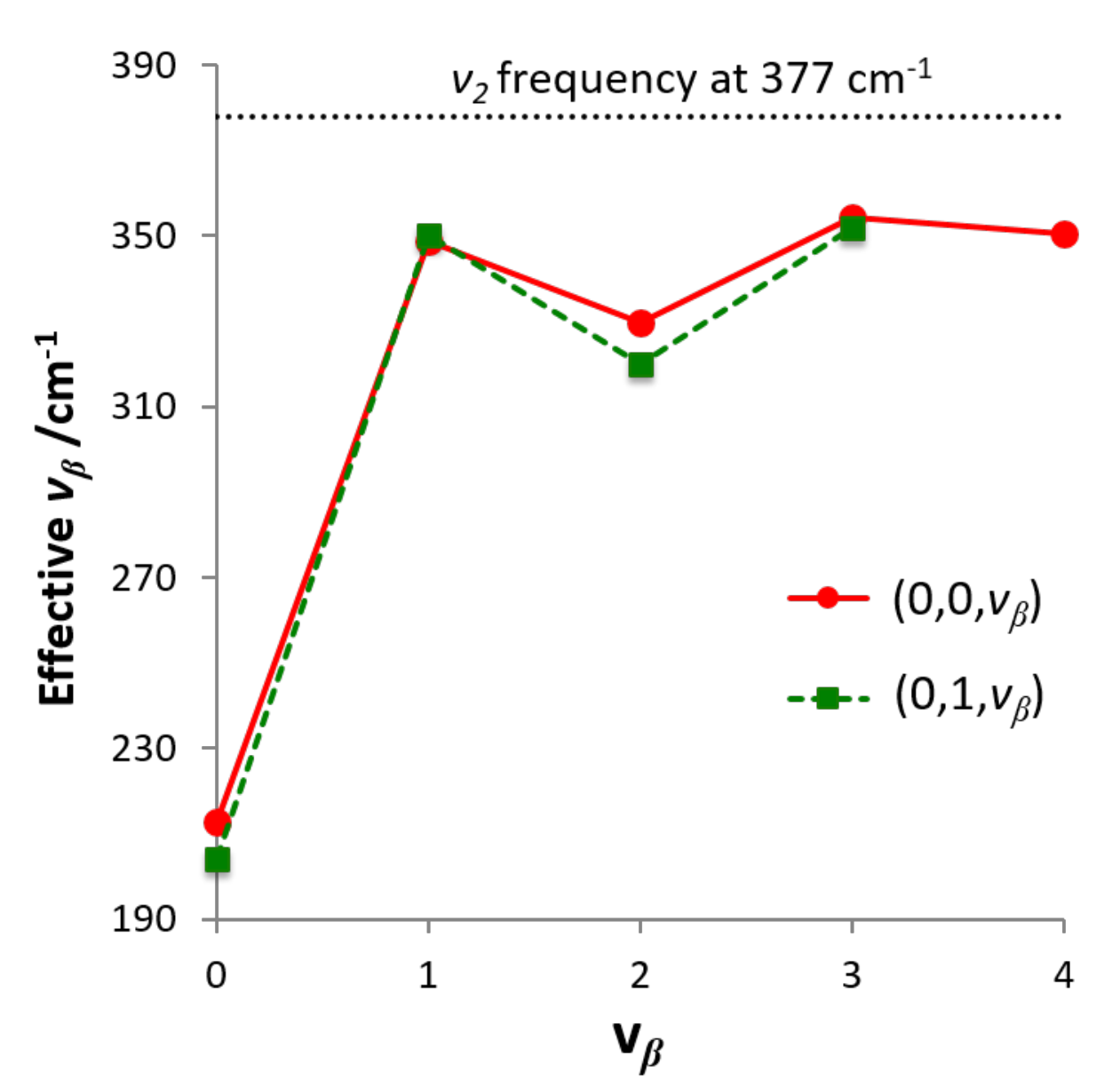}
\caption{The effective $\nu_\beta$ frequency in the $(0,0,v_\beta)$ and $(0,1,v_\beta)$ progressions. The dotted line at the top of the figure shows the approximate magnitude of the $\nu_2$ frequency.}
\label{fig:zigzag}
\end{figure}

\begin{figure}[b]
\includegraphics[width=3.1 in]{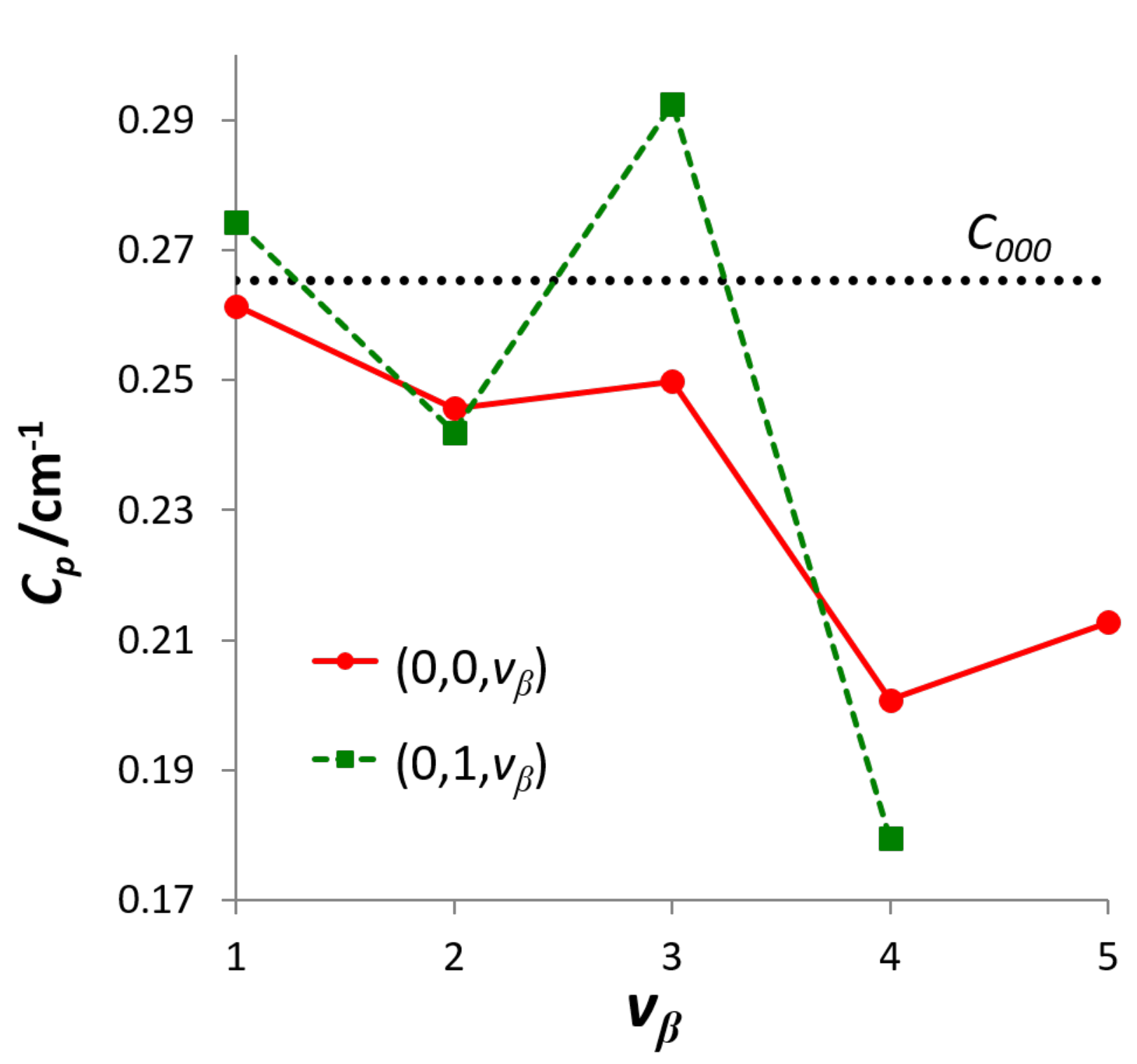}
\caption{Values of the experimental $perturbed$ $C$ rotational constants~\cite{{Yamanouchi1995},{Park2015SO2}} of levels in the $(0,0,v_\beta)$ and $(0,1,v_\beta)$ progressions, with $v_\beta>1$. The dashed line indicates the value of the $C$ constant of the zero-point level, $C_{000}$=0.2654 cm$^{-1}$.} 
\label{fig:C_progression}
\end{figure}

\subsection{The Zigzag Patterns in the $C$ Rotational Constants}
\label{subsec:Zigzag}

The alternating patterns in the Coriolis mixing angles shown in Fig.~\ref{fig:CorPolyad} are manifest as zigzag patterns in the $C_p$ rotational constants (the Coriolis-perturbed $C$ rotational constants). In Fig.~\ref{fig:C_progression}, values of the experimental effective $C_p$ constants~\cite{{Yamanouchi1995},{Park2015SO2}} of levels in the $(0,0,v_\beta)$ and $(0,1,v_\beta)$ progressions are plotted as a function of  $v_\beta$. It is evident that the $C_p$ rotational constants in both progressions follow a zigzag trend. The $C$ constant of the zero-point level ($C_{000}$) is used as a reference for the $C_p$ constants in Fig.~\ref{fig:C_progression}. Given that the Coriolis contribution to the $C$ constants outweighs the corrections from other contributions by an order of magnitude (with the exception of the $(0,0,1)_r$ level), the deviation of the  $C_p$ constant of a specific vibrational level from $C_{000}$ gives information about the mixing angle between that level and its Coriolis-interacting levels. 

The $C_p$ rotational constants of levels in the $(0,0,v_\beta)$ progression ($v_\beta\leq5$) are all smaller than $C_{000}$, since all of the Coriolis-interacting partners of those levels lie higher in energy. The zigzag pattern arises from oscillations in the magnitudes of the Coriolis mixing angles between the $(0,0,v_\beta)$ an $(0,1,v_\beta-1)$ levels as $v_\beta$ increases (see Fig.~\ref{fig:CorPolyad}). Note that the deviations of the $C_p$ constants from $C_{000}$ match the trend in the mixing angles between the $(0,0,v_\beta)$ and $(0,1,v_\beta-1)$ levels in Fig.~\ref{fig:CorPolyad}, i.e. a larger mixing angle leads to a larger deviation. The $C_p$ rotational constants of levels in the $(0,1,v_\beta)$ progression ($v_\beta\leq4$) oscillate around the value of $C_{000}$, because levels in this progression are affected by two competing Coriolis interactions with $(0,0,v_\beta+1)$ and $(0,2,v_\beta-1)$ levels. The two resulting mixing angles are not equal and their relative magnitudes alternate as $v_\beta$ increases (see Fig.~\ref{fig:CorPolyad}). Consequently, the $C_p$ constants of levels in the $(0,1,v_\beta)$ progression ($v_\beta\leq4$) oscillate around the value of $C_{000}$, as $v_\beta$ increases. 

We believe that a zigzag trend in the rotational constants of levels in a vibrational progression is a signature of a double-well on the PES. Similar observations have been made in other molecular systems with a small barrier on the PES~\cite{{Chan1960},{Duckett1976},{Duckett19762}}, although in other molecules, the rotational constants that exhibit the zigzag pattern are not strongly perturbed by Coriolis interactions, so the deviations of those rotational constants from the usual linear trend is about two or three orders of magnitude smaller than in the $C$ rotational constants of the $\tilde{\text{C}}$ state of SO$_2$. Thus, Coriolis interactions amplify the effect of a double-well structure on the rotational constants.  

\section{Conclusion}
\label{sec:conclusion}

In this work, an internal coordinate force field through quartic terms for the SO$_2$ $\tilde{\text{C}}$$^1$B$_2$ state has been derived. The force field fit incorporates vibrational and rotational information from two isotopologues of SO$_2$ (S$^{16}$O$_2$ and S$^{18}$O$_2$), and, in particular, it includes nine recently measured $b_2$ symmetry levels of S$^{16}$O$_2$~\cite{{Park2015},{Park2015SO2}} . The good agreement between the experimental and calculated values of observables, many of which are not directly included in the fit, indicates that the force field is physical and accurate. In particular, the Franck-Condon intensities and the Coriolis-perturbed effective $C$ rotational constants of highly anharmonic $\tilde{\text{C}}$ state vibrational levels are well reproduced using our force field. The force field, together with our recent direct observations of $b_2$ vibrational symmetry levels~\cite{{Park2015SO2},{Park2015}}, add crucial pieces of information to our understanding of the SO$_2$ $\tilde{\text{C}}$$^1$B$_2$ system. Key aspects of the dynamics predicted by the new force field are analyzed. 
 
The assumption of the separability of $\nu_2$ from the other two strongly Fermi-interacting modes breaks down for levels that lie $>$2000 cm$^{-1}$ above the $\tilde{\text{C}}$ state zero-point level. Franck-Condon interference effects, due to interactions among states that have different numbers of quanta of excitation in $\nu_2$, are found to be prevalent among the $\tilde{\text{C}}$ state vibrational levels above 2500 cm$^{-1}$. The presence of interference effects in this energy region invalidates vibrational assignment based on apparent $\nu_2$ progressions. However, using a two-step diagonalization procedure of the vibrational Hamiltonian, an unprecedented number of $\tilde{\text{C}}$ state vibrational levels can now be assigned. The vibrational levels are characterized in the Kellman basis, constructed explicitly via partial diagonalization of the Hamiltonian. Decoding the new classes of dynamics exhibited by levels above 2500 cm$^{-1}$, which are highly mixed even in the Kellman basis, poses an interesting challenge for future work.

The Coriolis interactions in the $\tilde{\text{C}}$ state are modeled in this work using second-order non-degenerate perturbation theory. Most importantly, we identify a rotational signature indicating the presence of the double-well structure of the PES. The anomalies in the $C$ rotational constants result from the staggering in the vibrational energy spacings, due to the double-well structure. 
  
\section*{AUTHOR INFORMATION}
\label{sec:author}
\subsection*{\bf{Corresponding Author}}
*Email: rwfield@mit.edu

\begin{acknowledgments}
The authors thank Professor Anthony Merer, Professor Michael Kellman, and Trevor Erickson for valuable discussions. This material is based upon work supported by the U.S. Department of Energy, Office of Science, Chemical Sciences, Geosciences, and Biosciences Division of the Basic Energy Sciences Office, under Award Number DE-FG02-87ER13671.
\end{acknowledgments}

\appendix

\section{Construction of the Hamiltonian Matrices}
\label{app:matrix}
Given a set of normal-mode force constants and the three parameters that describe a hump on the PES, a harmonic oscillator basis is used to construct a Hamiltonian matrix for each isotopologue (in our study, S$^{16}$O$_2$ and S$^{18}$O$_2$). The (rotationless) vibrational Hamiltonian matrix, as given in Eq.~(\ref{eq:Hamiltonian}), is block-diagonalized into $a_1$ and a $b_2$ symmetry blocks. For the purpose of fitting to the observed energy levels, a dimension of around 1800 basis states for each symmetry block is found to be sufficient for convergence of the eigenvalues. Harmonic oscillator basis states that have diagonal matrix elements smaller than 21,000 cm$^{\text{-1}}$ are included in the matrix. However, basis states with more than 22 quanta in $\nu_2$ are excluded from the matrix because they are found to be unnecessary for eigenvalue convergence. Each Hamiltonian matrix (two for each isotopologue) is then diagonalized to obtain both eigenvalues and eigenvectors. Note that the matrix size necessary for eigenvalue convergence is large, despite the fact that only the lowest 60 eigenstates of each Hamiltonian are studied in the current work. The large size of the basis set is required because the harmonic basis set is not the most physical or efficient representation of the the $\tilde{\text{C}}$ state molecular system, primarily due to the presence of the double well structure of the PES.

\section{Semi-classical analysis of the Fermi polyads}
\label{app:kellman}

Based on visual inspection of the wavefunction shapes, the Fermi polyads of interest in Fig.~\ref{fig:2Dpolyad} seem to exhibit semiclassical dynamics that correspond to the dynamics in the Zone III on the catastrophe map used in Kellman's work~\cite{{Kellman1990cat},{Kellman1990new}}. The wavefunction of the highest member of each polyad for the $\tilde{\text{C}}$ state of SO$_2$ (see Fig.~\ref{fig:2Dpolyad}) resembles the wavefunction assigned as [0,14]$_{\text{IIIbc}}$ in the first subfigure of Fig.~\ref{fig:zone3} (albeit with opposite pointing direction), while wavefunctions of the lower members of each polyad in Fig.~\ref{fig:2Dpolyad} resemble the rest of the wavefunctions in Fig.~\ref{fig:zone3}. Therefore, the lower members of each Fermi polyad in the $\tilde{\text{C}}$ state undergo what Kellman calls a resonance collective motion. The wavefunctions of those levels open up $completely$ in the positive $q_1$ direction. The highest member of each polyad, $(v_\alpha,0,0)_l$, is a mixture of normal-mode and resonance collective mode states. The wavefunction of each highest-energy polyad level has contributions both from resonant collective motions, with the wavefunction opening up in the negative $q_1$ direction, and from normal-modes, which prevents the wavefunction from $completely$ opening up, in contrast to the shapes of the wavefunctions of the lower polyad members. 

\begin{figure}[b]
\includegraphics[width=4 in]{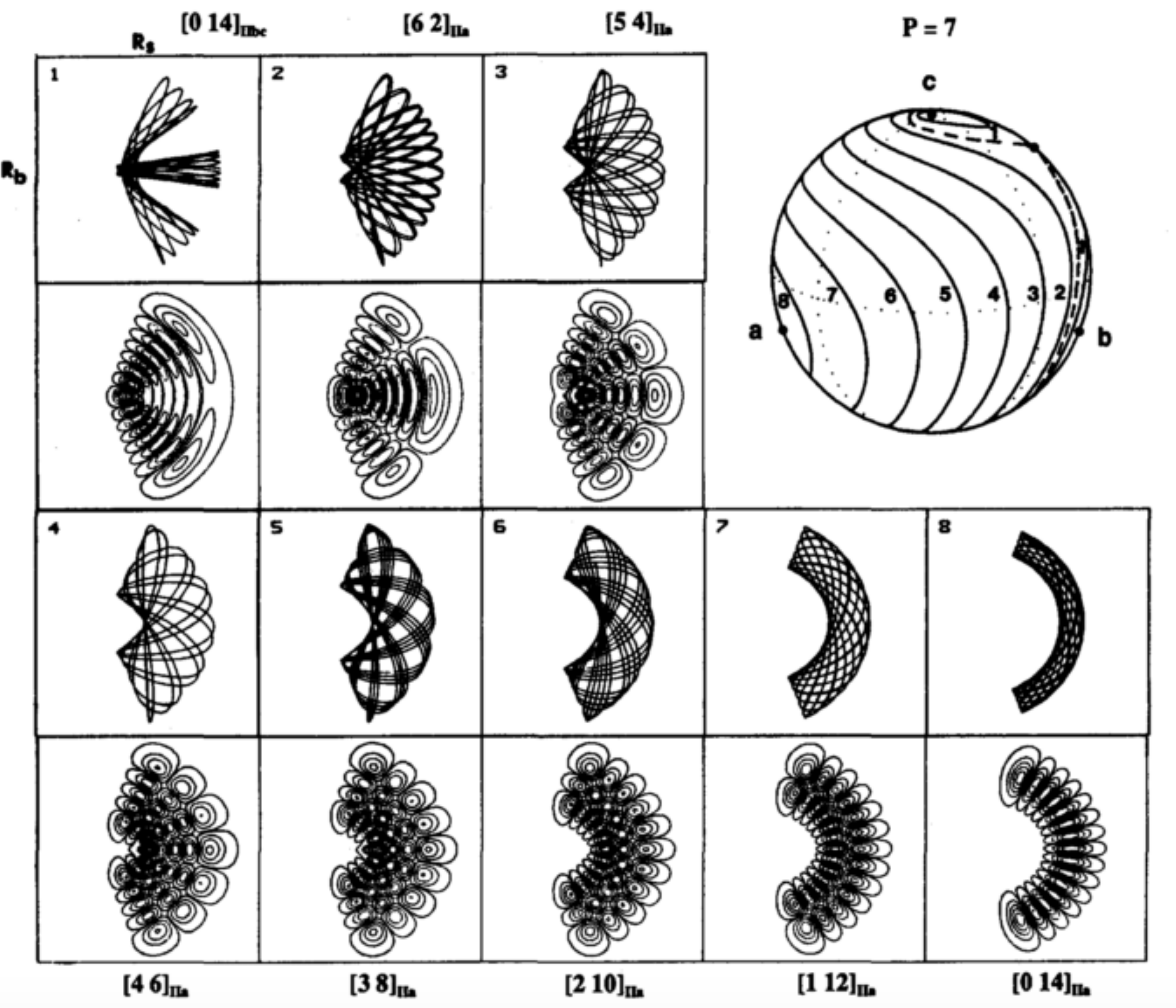}
\caption{Semiclassical dynamics in Zone III of Kellman's catastrophe map. Reproduced with permission from J. Chem. Phys. $\textbf{93}$, 5821 (1990). Copyright 1990 American Institute of Physics. Note that there are typographical errors in the assignments given in this figure. All of the `II' in the subscripts of the assignment labels should be replaced by `III'.}
\label{fig:zone3}
\end{figure}

Note that the majority of states within a given polyad in Zone III lie in region $a$ of the classical polyad phase sphere (hence the subscripts `IIIa' in the assignments of those states in Fig.~\ref{fig:zone3}), where semiclassical trajectories of levels in this region correspond to a resonant collective mode. In the molecular systems Kellman has studied~\cite{Kellman1990new}, the [0,14]$_{\text{IIIbc}}$ level in Fig.~\ref{fig:zone3} spans both regions $b$ and $c$. Trajectories in region $b$ correspond to a different resonant collective mode (the wavefunction opens up in the direction opposite to the first resonant collective mode), while trajectories in region $c$ correspond to normal stretching mode motion. As a result, the [0,14]$_{\text{IIIbc}}$ wavefunction has characters of both resonant collective mode and normal-mode wavefunctions. 

We emphasize that, while we are using Kellman's assignment scheme, our notations are slightly different. First, in Kellman's notation, the subscripts indicate both the zone on the catastrophe map in which the polyad lies and the region on the polyad phase sphere where the level in question is located. In our notation, this semiclassical information is not included; instead, we give an $r$ or $l$ label, which indicates the direction in which the wavefunction opens. Our intention is to help the reader visualize the shape of the wavefunction. Second, our notations for the assignments of the highest member of the polyad differ from Kellman's choice. In his notation, the level we assign as (3,0,0)$_l$ would be (0,0,6)$_{\text{IIIbc}}$. Aside from the difference in the subscripts just discussed, the definitions of our quantum numbers differ from those used by Kellman~\cite{Kellman1990new}. In Kellman's choice, the quantum numbers for levels that span both regions $b$ and $c$ are defined with respect to the fixed point $b$ in the polyad phase sphere shown in Fig.~\ref{fig:2Dpolyad}~\cite{Kellman1990new}, while in our choice, the quantum numbers for the same state are defined with respect to the normal-mode fixed point $c$, which, in the specific case of the $\tilde{\text{C}}$ state of SO$_2$, the fixed point $c$ corresponds to the symmetric stretching normal-mode. Those two notations should be equally valid, since trajectories in regions $b$ and $c$ are degenerate, as long as one keeps in mind that semiclassically, the $(v_\alpha,0,0)_l$ state is a mixture of states in both regions $b$ and $c$. 

\bibliographystyle{aipauth4-1}
\newpage

\bibliography{SO2_paper}



\end{document}